\documentclass[onecolumn,amsmath,amssymb]{revtex4}

\usepackage{graphicx}
\usepackage{dcolumn}
\usepackage{bm}
\usepackage{subfigure}
\usepackage{textcomp}
\usepackage{threeparttable}
\usepackage{booktabs}
\usepackage{slashed}
\usepackage{graphicx}
\usepackage[dvips]{color}
\usepackage{type1cm}
\renewcommand{\TPTtagStyle}%
{\normalsize\textit}

\begin{document}
\fontsize{12pt}{12pt}\selectfont

\title{Chiral Hall Effect and Chiral Electric Waves}
\author{Shi Pu$^{1,2,3}$\footnote{pushi@ntu.edu.tw}, Shang-Yu Wu$^{4,5,6}$\footnote{loganwu@gmail.com}, Di-Lun Yang$^{7,8}$\footnote{dy29@phy.duke.edu}}
\affiliation{$^1$Department of Physics, National Center for Theoretical Sciences,
and Leung Center for Cosmology and Particle Astrophysics,
National Taiwan University, Taipei 10617, Taiwan\\
$^2$Interdisciplinary Center for Theoretical Study and Department of Modern Physics,
University of Science and Technology of China, Hefei 230026, China\\
$^3$Institute for Theoretical Physics, Goethe University, Max-von-Laue-Str. 1, 60438 Frankfurt am Main, Germany\\
$^4$Institute of physics, National Chiao Tung University, Hsinchu 300, Taiwan.\\
$^5$National Center for Theoretical Science, Hsinchu, Taiwan.\\
$^6$Yau Shing Tung Center, National Chiao Tung University, Hsinchu, Taiwan.\\
$^7$Department of Physics, Duke University, Durham, North Carolina 27708, USA.\\
$^8$Department of Physics, Chung-Yuan Christian University (CYCU), Chung-Li 32023, Taiwan.}
\date{\today}
\begin{abstract}
We investigate the vector and axial currents induced by external electromagnetic fields and chemical potentials in chiral systems at finite temperature. Similar to the normal Hall effect, we find that an axial Hall current is generated in the presence of the electromagnetic fields along with an axial chemical potential, which may be dubbed as the "chiral Hall effect"(CHE). The CHE is related to the interactions of chiral fermions and exists with the a nonzero axial chemical potential. We argue that the CHE could lead to nontrivial charge distributions at different rapidity in asymmetric heavy ion collisions.   
Moreover, we study the chiral electric waves(CEW) led by the fluctuations of the vector and axial chemical potentials along with the chiral electric separation effect(CESE), where a density wave propagates along the applied electric field. Combining with the normal/chiral Hall effects, the fluctuations of chemical potentials thus result in Hall density waves. The Hall density waves may survive even at zero chemical potentials and become non-dissipative. We further study the transport coefficients including the Hall conductivities, damping times, wave velocities, and diffusion constants of CEW in a strongly coupled plasma via the AdS/CFT correspondence.        
\end{abstract}
\maketitle
\section{Introduction}
The anomalous transport induced by electromagnetic fields has been widely studied recently. In the presence of an axial chemical potential, a vector current will propagate parallel to an applied magnetic field led by triangle anomalies, which is the renowned chiral magnetic effect(CME)\cite{Kharzeev:2007tn,Kharzeev:2007jp,Kharzeev:2010gd,Son:2004tq}. Although this effect was initially found in the deconfined phase, it may exist in the hadronic phase as well\cite{Asakawa:2010bu}. Analogous to CME, a vector chemical potential can generate an axial current along the magnetic field, which is the so-called chiral separation effect(CSE)\cite{Fukushima:2008xe}.
These effects have been further derived from varieties 
of different approaches, including 
relativistic hydrodynamics \cite{Son:2009tf,Pu:2010as,Sadofyev:2010pr,Kharzeev:2011ds,Nair:2011mk},
kinetic theory \cite{Gao:2012ix,Son:2012wh,Stephanov:2012ki,Son:2012zy,Chen:2012ca,Pu:2012wn,Chen:2013iga,Manuel:2013zaa,Manuel:2014dza,Satow:2014lva,Duval:2014ppa},
and lattice simulations\cite{Abramczyk:2009gb,Buividovich:2009wi,Buividovich:2010tn,Yamamoto:2011gk,Bali:2014vja}.
Also they were analyzed in the strongly coupled plasmas through the AdS/CFT correspondences\cite{Yee:2009vw,Rebhan:2009vc,Gorsky:2010xu,Gynther:2010ed,Kalaydzhyan:2011vx,Hoyos:2011us,Gahramanov:2012wz}.
However, in the Sakai-Sugimoto(SS) model as a commonly used model for AdS/QCD\cite{Sakai:2004cn,Sakai:2005yt}, CME may disappear when requiring both gauge-invariance and conservation of the vector current\cite{Rebhan:2009vc,Yee:2009vw,Gynther:2010ed,Rubakov:2010qi}. 
For a recent review of CME/CSE and related topics, see e.g. \cite{Kharzeev:2012ph,Liao:2014ava} and the references therein. 
The effects are particularly important in the heavy ion experiments, where the charge separation could arise from the strong magnetic field produced from the colliding nuclei and non-vanishing chemical potentials in the quark gluon plasma(QGP). In light of CME/CSE, it was proposed that the thermal fluctuations of the vector and axial chemical potentials in thermal plasmas can further result in density waves propagating along the magnetic field as the chiral magnetic waves(CMW)\cite{Kharzeev:2010gd}. In \cite{Kharzeev:2010gd}, the dispersion relation of CMW was investigated in the framework of the SS model with zero chemical potentials. As shown in \cite{Burnier:2011bf}, the CMW could generate a chiral dipole and a charge quadrapole in QGP, which may contribute to the charge asymmetry of elliptic flow $v_2$ measured in the relativistic heavy ion collider(RHIC)\cite{Wang:2012qs,Ke:2012qb}. Further study of CMW in an expanding QGP can be found in \cite{Taghavi:2013ena}. In addition to the anomalous effects, the strong magnetic field also gives rise to profound phenomena such as the enhanced photon production \cite{Basar:2012bp,Fukushima:2012fg,Bzdak:2012fr,Wu:2013qja,Yee:2013qma,Muller:2013ila},
which could be crucial for the large elliptic flow observed in RHIC \cite{Adare:2011zr} and in the large hadron collider(LHC)\cite{Lohner:2012ct},
the production of heavy quarkonia\cite{Yang:2011cz,Machado:2013rta,Alford:2013jva}, and the modified shear viscosity of QGP\cite{Nam:2013fpa,Critelli:2014kra}.

In addition to the strong magnetic field, a strong electric field could be produced in heavy ion collisions as well. The strong electric field having the magnitude of $m_{\pi}^2$ with $m_{\pi}$ being the mass of pions could exist in the asymmetric collisions such as the Au nucleus to the Cu nucleus in early times\cite{Hirono:2012rt}. 
Furthermore, the electric field can be comparative to that of the magnetic field on the basis of event-by-event fluctuations even in the symmetric collisions\cite{Bzdak:2011yy,Deng:2012pc}. A novel phenomenon called chiral electric effect(CESE) has been proposed in \cite{Huang:2013iia}, where an axial current can be produced parallel to the electric field in the presence of both vector and axial chemical potentials. The direct-current(DC) conductivity of the axial charge was found to be proportional to the product of the axial chemical potential and the vector chemical potential in the weakly coupled QED with small chemical potentials compared to the temperature of the medium. Such a relation was later verified in the strongly coupled scenario in the SS model\cite{Pu:2014cwa}. Moreover, the relation is approximately hold even for large chemical potentials. Unlike CME/CSE, since CESE is not contributed by the Chern-Simons(CS) term related to the axial anomaly but only by the nonzero vector and axial chemical potentials, the axial current from CESE in the SS model is well defined. Besides, in Ref.\cite{Chen:2013tra}, the studies of electric conductivities of
non-singlet currents in a weakly coupled QCD system with multi-flavors
implies that the similar behavior of axial conductivities in small
chemical potentials could also observed in QCD. Similar to CMW, the density fluctuations may induce the propagating waves along the electric field as the chiral electric waves(CEW)\cite{Huang:2013iia}. In phenomenology, the combination of CME and CESE could possibly generate quadrapole distribution of charge particles when the electric field and magnetic field are perpendicular to each other as in the asymmetric collisions. It is thus imperative to further investigate CESE and CEW. 

We will continue our study in \cite{Pu:2014cwa} to further explore the CESE and CEW with arbitrary chemical potentials. From the classical electrodynamics, the presence of both an electric field and a magnetic field perpendicular to each other should yield a Hall current perpendicular to both applied fields. Since the CESE is analogous to the normal transport process which is governed by the interaction between the chiral particles, we will find an axial Hall current similar to the axial current parallel to the electric field in the absence of the axial anomaly. 

In general, we analyze the CESE, classical Hall effect, and chiral Hall effect(CHE) in chiral systems in the presence of external electromagnetic fields and also investigate the propagating waves caused by the density fluctuations with arbitrary chemical potentials. Nevertheless, we will assume that the interaction between the chiral particles dominates the topological effect and thus neglect the CME/CSE. In addition, we will implement the SS model to compute the transport coefficients including the damping times, wave velocities, and diffusion constants of CEW. 

For convenience, we briefly summarize CME, CSE, CESE,
CHE, CMW and CEW in Tab. \ref{tab:A-brief-summary}.

This paper is organized in the following order. In section \ref{Hall_effects}, we review the classical Hall effect and derive the axial Hall current. In section \ref{ph_implications}, we will discuss the phenomenological implications of the CESE and CHE. In section \ref{density_waves}, we then generalize both the CMW and CEW to the cases with arbitrary chemical potentials. Also, we analyze the CEW on the basis of the CESE and CHE. In section \ref{SS_model}, we review the setup of the SS model in a chiral symmetric phase at finite temperature with chemical potentials and a constant electric field perpendicular to a constant magnetic field, where we further derive the axial Hall current. In section \ref{CHE_holography}, we will analyze the CESE and CHE in different limits and present the numerical results in the framework of the SS model. In section \ref{CEW_holography}, we numerically solve for CEW in the SS model. In addition, we briefly compare the CEW at small chemical potentials in the strongly coupled QCD with that in the weakly coupled QED. Finally, we make a brief summary and outlook in section \ref{sum_outlook}. Throughout the paper, we will set ${\bf B}=B_x\hat{x}$, ${\bf E}=E_y\hat{y}$ when we discuss the Hall and chiral Hall effects, where ${\bf E}$ and ${\bf B}$ denote the external electric and magnetic fields in our systems.

\begin{table}
\caption{A brief summary to CME, CSE, CESE, CHE, CMW and CEW. Here $\mu_{V},\mu_{A}$
are vector and axial vector chemical potentials, respectively. $\mathbf{j}_{v}$
and $\mathbf{j}_{a}$ are vector and axial vector current. $\sigma_{a},(\sigma_{v})_{zy},(\sigma_{a})_{zy}$
are transport coefficients. \label{tab:A-brief-summary} }

\centering{}%
\begin{tabular}{|c|c|c|}
\hline 
 & Currents & Possible phenomena\tabularnewline
\hline 
\hline 
Chiral Magnetic Effect & $\mathbf{j}_{v}=\frac{e}{2\pi^{2}}\mu_{A}\mathbf{B},$ & charge separation along $\mathbf{B}$ field \tabularnewline
\hline 
Chiral Separation Effect & $\mathbf{j}_{a}=\frac{e}{2\pi^{2}}\mu_{V}\mathbf{B},$ & chirality separation along $\mathbf{B}$ field \tabularnewline
\hline 
Chiral Electric Separation & $\mathbf{j}_{a}=\sigma_{a}\mathbf{E},$  & charge and chirality separation \tabularnewline
Effect &  & along $\mathbf{E}$ field\tabularnewline
\hline 
Chiral Hall Effect & $j_{v,z}=(\sigma_{v})_{zy}E_{y},$ & charge and chirality separation\tabularnewline
 & $j_{a,z}=(\sigma_{a})_{zy}E_{y},$ & in rapidity direction\tabularnewline
\hline 
\hline 
Chiral Magnetic Wave & Evolution equations for  & density wave induced by magnetic field\tabularnewline
 & currents with CME, CSE & and charge separation along $\mathbf{B}$ field\tabularnewline
\hline 
Chiral Electric Wave  & Evolution equations for  & density wave induced by electric field,\tabularnewline
 & currents with CESE, CHE & charge separation along $\mathbf{E}$ field\tabularnewline
 &  & and rapidity direction\tabularnewline
\hline 
\end{tabular}
\end{table}

\section{Hall effect and chiral Hall effect}\label{Hall_effects}
In classical physics, the Hall current is coming from the balance
of two forces in a conductor, i.e. the electric and magnetic forces,
\begin{equation}
e\mathbf{E}=-e\mathbf{v}\times\mathbf{B},\label{eq:Lorentz_force_01}
\end{equation}
where $\mathbf{v}$ is the velocity of a single electron or positron and
$e$ is the charge of particles. In a many body system, multiplying
the number density of particle, $n$, to the both sides of above equations,
yields, 
\begin{equation}
ne\mathbf{E}=-ne\mathbf{v}\times\mathbf{B}.
\end{equation}

Recalling the charge currents in an equilibrium state, $j_{eq0}=n$,
$\mathbf{j}_{eq}(x)=n\mathbf{\bar{v}}$, with $\bar{\mathbf{v}}$
the average of the particles' velocities at point $x$. Without external
fields, the system will be homogenous and $\mathbf{j}_{eq}(x)=n\mathbf{\bar{v}}\rightarrow0$
in the local rest frame. In the present of external fields, most of
particles will be accelerated by $\mathbf{E}$ field and become
the normal electric conducting flow. While a few particles, which move
orthogonal to $\mathbf{E},\mathbf{B}$ fields and satisfy Eq. (\ref{eq:Lorentz_force_01}),
will not feel the external fields and cause a new current $\mathbf{j}$.
Neglecting high order terms of \textbf{$\mathbf{E},\mathbf{B}$},
this new current will satisfy, 
\begin{equation}
j_{0}\mathbf{E}=-\mathbf{j}\times\mathbf{B},
\end{equation}
Since the current is proportional to the absolute value of \textbf{$\mathbf{E}$}
field, one can consider it as another conducting flow and introduce
the conductivity tensor as, 
\begin{equation}
j_{i}=\sigma_{ij}eE_{j},
\end{equation}
If $\mathbf{E}=E\hat{y}$, $\mathbf{B}=B\hat{x}$, then we find, 
\begin{equation}
\sigma_{zy}=-\frac{n}{eB},\label{eq:Hall_strong_B_01}
\end{equation}
which is Hall conductivity. Note that, above discussion cannot be
applied to a small $\mathbf{B}$ field case, otherwise, the balance
of two forces will never be reached, if $|\mathbf{E}|>c|\mathbf{B}|,$
with $c$ the speed of light. Since, if $B=0$, there will be no Hall
effect, therefore, we expect that in small $\mathbf{B}$ case, the
Hall conductivity will be, 
\begin{equation}
\sigma_{zy}=-n\tau_{H}^{2}eB,\label{eq:Hall_weak_B_01}
\end{equation}
where $\tau_{H}$ is parameter with dimension $MeV^{-2}$. Physically,
$\tau_{H}$ is related to the interaction between particles. Since
when $\mathbf{B}$ is too weak, the interaction from particles will
give an effective force to each particles and the force will help
to satisfy Eq. (\ref{eq:Lorentz_force_01}). As shown in Eq.(\ref{eq:sol_Langevin_01})
at Appendix A, the $\tau_{H}$ can be solved in weakly magnetic field
limit in Langevin equations (\ref{lorentzf}), i.e. $\tau_{H}=\xi M$
, with $\xi$ the drag coefficient related to the interactions and
$M$ the mass of particles. A systematic discussion in both strong
and weak $B$ limit via Langevin equation and Boltzmann equation with
relaxation time approaches is shown in Appendix A. 

Although it seems that the normal electric conductivities $\sigma_{ii}$
vanishes in this discussion, for a fixing \textbf{$\mathbf{E}$} and
$\mathbf{B}$ fields, as we mentioned, only a few particles could
satisfy Eq. (\ref{eq:Lorentz_force_01}) and others will still be
accelerated by the \textbf{$\mathbf{E}$} field. Therefore, the normal
electric conducting flow is still there. This can be understood in
the language of the Lagevin equations or Boltzmann equations, as shown in Eq. Appendix A.

Now let us extend our discussion to a chiral fermion system. In this
case, the single charge current will become the right and left handed
currents, $\mathbf{j}_{R}$ and $\mathbf{j}_{L}$. In the present
of axial chemical potential $\mu_{A}$, the Hall conductivities in
Eq. (\ref{eq:Hall_strong_B_01}, \ref{eq:Hall_weak_B_01}) for $\mathbf{j}_{R/L}$
will be different because of $n_{R}\neq n_{L}$,
\begin{equation}
(j_{R/L})_i=(\sigma_{R/L})_{ij}E_{j}.
\end{equation}
Therefore, the vector and axial vector currents are defined as, 
\[
\mathbf{j}_{v}=\frac{1}{2}(\mathbf{j}_{R}+\mathbf{j}_{L}),\;\mathbf{j}_{a}=\frac{1}{2}(\mathbf{j}_{R}-\mathbf{j}_{L}).
\]
There will be a chiral Hall effect(CHE) caused by the differences of Hall
conductivities of right and left handed fermions. If $\mathbf{E}=E_y\hat{y}$,
$\mathbf{B}=B_x\hat{x}$, we can define the normal Hall conductivity,
\begin{equation}
(\sigma_{v})_{zy}=-(\sigma_v)_{yz}=\frac{1}{2}(\sigma_{R}+\sigma_{L})_{zy},
\end{equation}
and the chiral Hall conductivity,
\begin{equation}
(\sigma_a)_{zy}=-(\sigma_a)_{yz}=\frac{1}{2}(\sigma_{R}-\sigma_{L})_{zy}.
\end{equation}

Now we can discuss the property of the normal and chiral Hall conductivity. The parity transformation, $\mathbf{x\rightarrow-x}$, will lead to
\begin{equation}
(\sigma_{a})_{zy}(\mathbf{x})=-(\sigma_{a})_{zy}(-\mathbf{x}),\;(\sigma_{v})_{zy}(\mathbf{x})=(\sigma_{v})_{zy}(-\mathbf{x}),
\end{equation}
which implies that $\sigma_{5H}\propto\mu_{A}$, since in the macroscopic
scaling, there is only a pseudo scalar in our system, $\mu_{A}$.
In a small $\mu_{V}$ and $\mu_{A}$ limit, from Eq. (\ref{eq:Hall_strong_B_01},
\ref{eq:Hall_weak_B_01}), we find, in a weak $\mathbf{B}$ field
case, 
\begin{eqnarray}\label{paritysmallmu}
(\sigma_{v})_{zy} & = & \chi_{e}eB_x\mu_V,\nonumber \\
(\sigma_{a})_{zy} & = & \chi_{5e}eB_x\mu_A,\label{eq:power_counting_Hall_weak_01}
\end{eqnarray}
and in a strong $\mathbf{B}$ field case, 

\begin{eqnarray}
(\sigma_{v})_{zy} & = & \chi_{e}^{\prime}T^{2}\mu_V/(eB_x),\nonumber \\
(\sigma_{a})_{zy} & = & \chi_{5e}^{\prime}T^{2}\mu_A/(eB_x),\label{eq:power_counting_hall_strong_02}
\end{eqnarray}
with $\chi_{e,5e},\chi_{e,5e}^{\prime}$ dimensionless function of
$T$ and $\mathbf{E}$. 

A similar effect can be observed in an anisotropic fluid with Berry
phase. When neglecting the interactions between particles, at an external
electric and magnetic fields, the effective velocity of a single right
handed Weyl fermions reads \cite{Son:2012zy,Stephanov:2012ki,Chen:2012ca},
\begin{equation}
\dot{\mathbf{x}}=\frac{\mathbf{p}}{|\mathbf{p}|}+\mathbf{E}\times\boldsymbol{\Omega}+\mathbf{B}(\frac{\mathbf{p}}{|\mathbf{p}|}\cdot\boldsymbol{\Omega}),
\end{equation}
where $\mathbf{p}$ is the momentum of that particle and $\boldsymbol{\Omega}=\mathbf{p}/(2|\mathbf{p}|^{3})$
is the Berry curvature. The right handed current is defined by 
\begin{equation}
\mathbf{j}_{R}=\int\frac{d^{3}p}{(2\pi)^{3}}\dot{\mathbf{x}}f(x,p)=n_{R}\mathbf{v}+\mathbf{E}\times\int\frac{d^{3}p}{(2\pi)^{3}}\boldsymbol{\Omega}f(x,p)+\frac{\lambda}{2}\mu_{A}\mathbf{B},
\end{equation}
where $f(x,p)$ is the distribution function. The third term gives
the CME. Once $f(x,p)$ is anisotropic in momentum space, the second
term will induce a current perpendicular to the electric field. However,
this current can survive even if \textbf{$\mathbf{B}=0$}.
In a 2+1 dimensional non-interacting fermion system, similar effects
from Chern-Simions term in an effective action of 2+1 dimensional
QED are also appear \cite{Chen:2013dca}. 
Quite different with above effects, the Hall and chiral Hall effects
depends on interactions and can survive without topological effects
and Berry phase.  

\section{Phenomenological Implications}\label{ph_implications}
The CESE and CHE may have important implications for the phenomenology of heavy ion collisions. For simplicity, we consider a system with only $u$ and $\bar{u}$
quarks in the following discussion. If $\mu_{V}>0$ or $\mu_{V}<0$,
there will be more particles or anti-particles, respectively. On the contrary,
if $\mu_{A}>0$ or $\mu_{A}<0$, there are more right or left handed
fermions. 

In the following discussion, we will assume there is a small net positive
$\mu_{V}$ after the two nuclei collide with each other since totally
there are more particles than anti-particles. For CME and CSE, a finite
$\mu_{A}$ is not necessary, since the CSE will induce a finite $\mu_{A}$
with the evolution. Nevertheless, to simplify the condition in the presence of 
both electric and magnetic fields, we ignore the detail of the axial charge distribution 
from CSE and just assume there 
exists a net positive $\mu_{A}$
as an initial condition when we discuss CESE and CHE. One can consider
the net $\mu_{A}$ to be induced by CSE
or by fluctuations or topological
transitions of QCD vacuum in each events. 

\begin{figure}
\begin{minipage}[t]{1\columnwidth}%
\subfigure[CSE and CME \label{fig:CSE-and-CME}]{\includegraphics[scale=0.42]{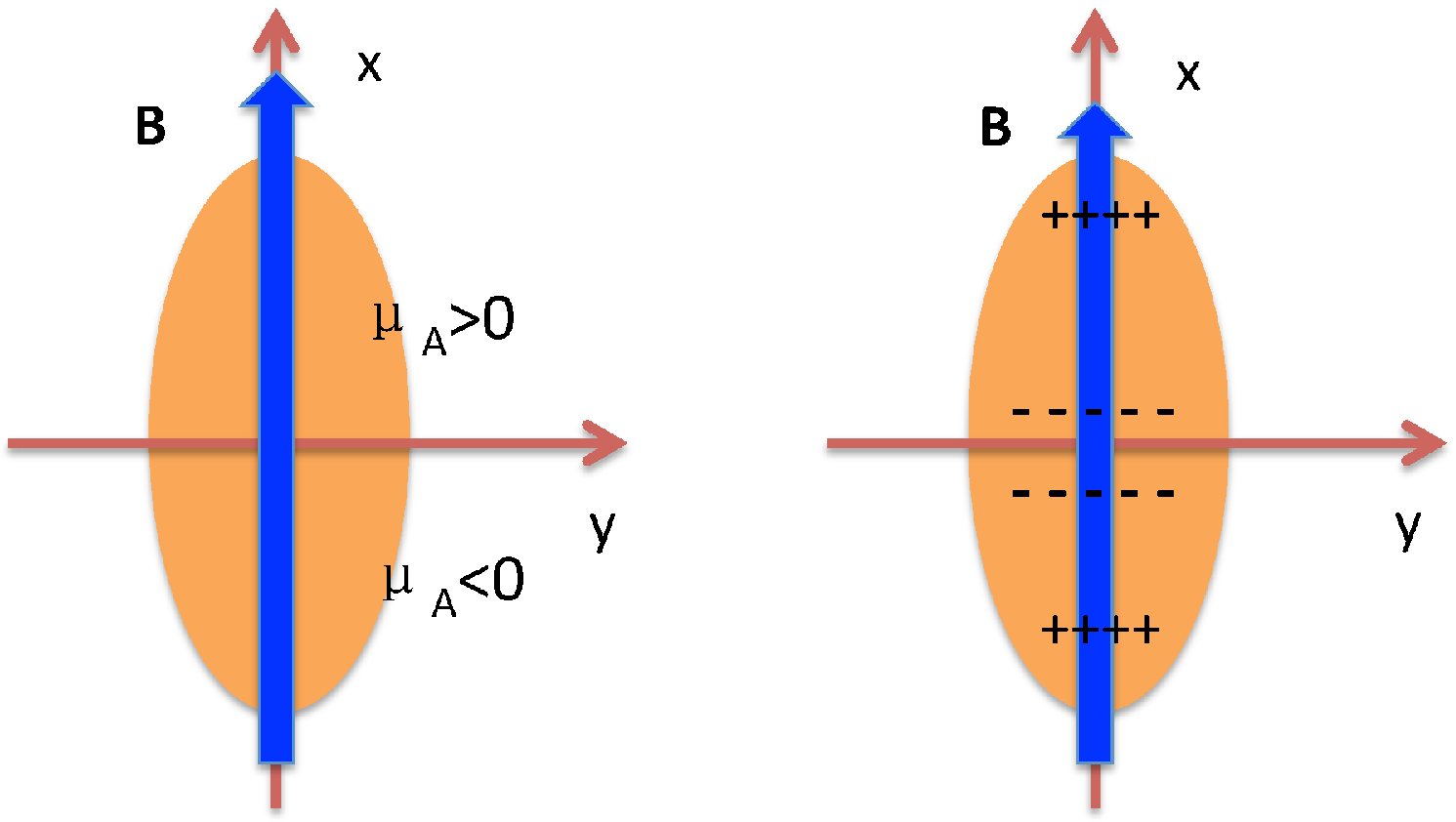}
}
\hspace {1cm}
\subfigure[CESE and CME \label{fig:CESE-and-CME}]{\includegraphics[scale=0.42]{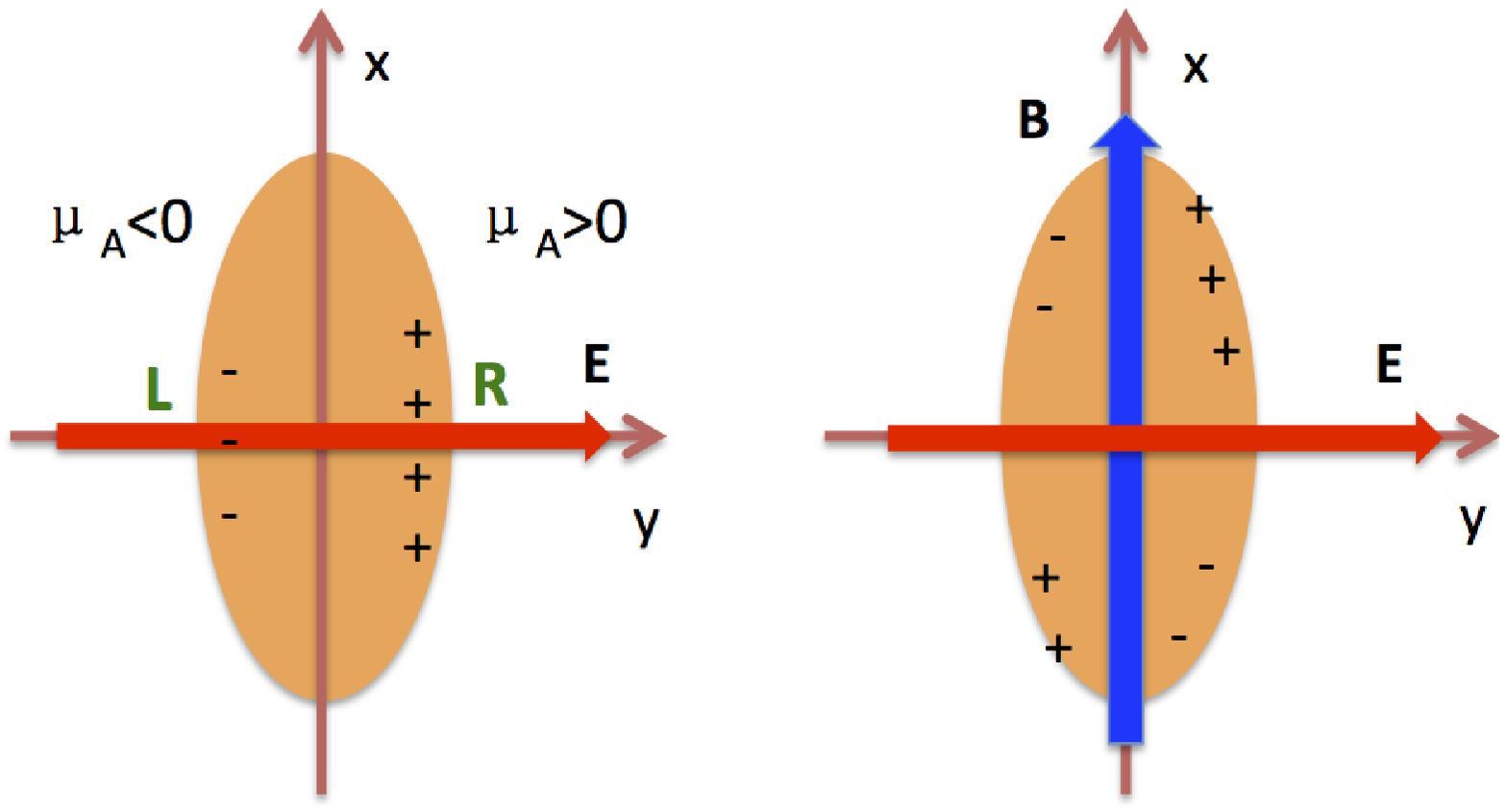}
}%
\end{minipage}
\par
\begin{centering}
\subfigure[Hall and chiral Hall effects\label{fig:Hall-and-chiral}]{\includegraphics[scale=0.42]{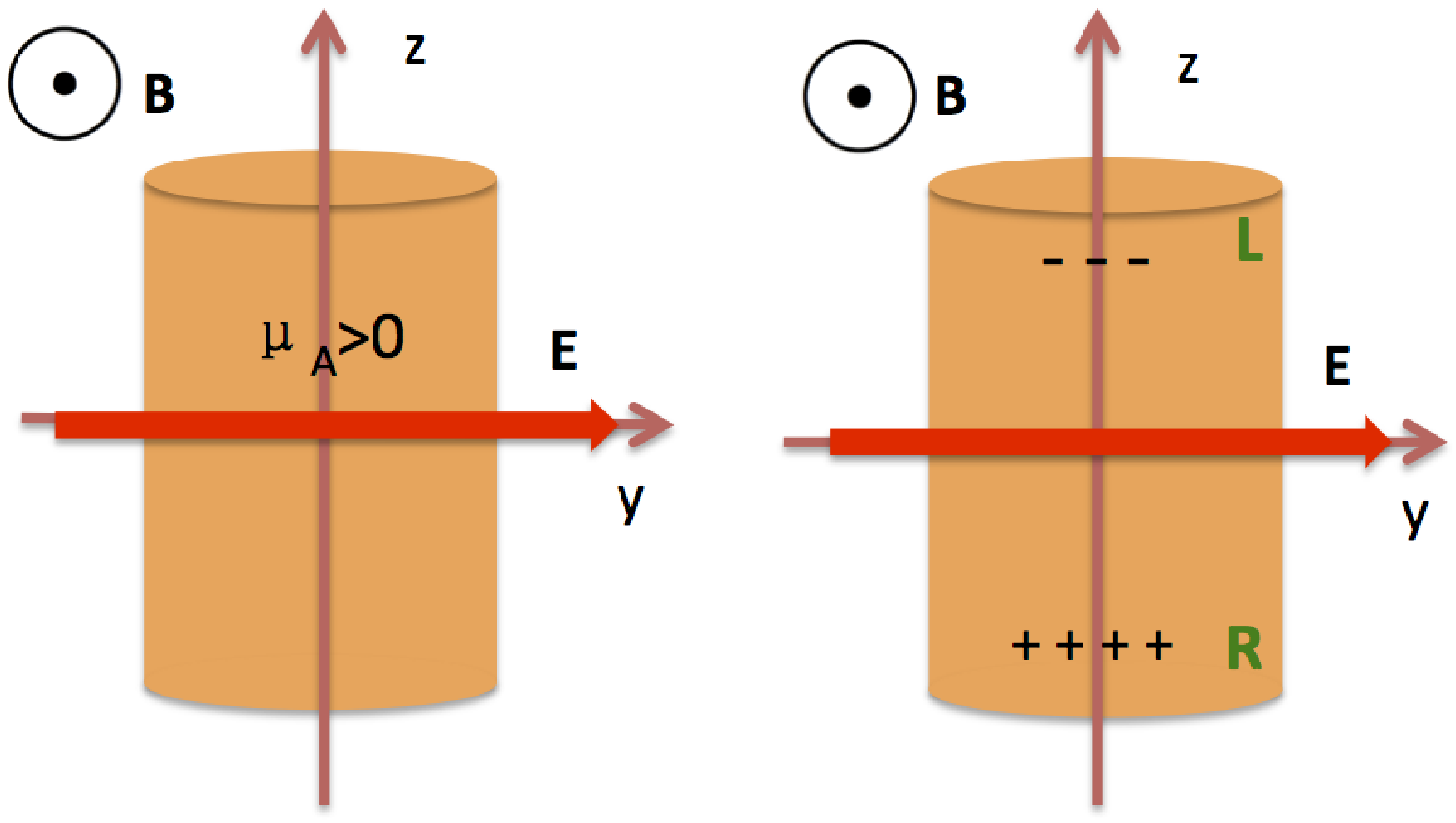}
}
\par\end{centering}
\caption{A schematic illustration for (a) CSE, (b) CESE and (c) Hall and chiral
Hall effects. In (b,c), for simplicity, we have assumed the system
has a $\mu_{A}>0$. In those figures, two nuclei collide through the $z$
direction. The strong magnetic and electric fields are at $x$ and
$y$ directions. The origin of the frame is set to be the center
of the fireball. In (c), we find a possible charge and chirality separation
induced by Hall and chiral Hall effects in the $z$ direction. \label{fig:cartoon}}
\end{figure}

Firstly, we will give a brief review to the scenario caused by the CME and CSE. In the
relativistic non-central heavy ion collisions, two nuclei collide with
each other through the $z$ direction as the beam direction shown in Fig. \ref{fig:cartoon}
and a very strong magnetic field $\mathbf{B}$ appears perpendicular
to the reaction plane, which is at the $x$ direction in Fig. \ref{fig:CSE-and-CME}.
According to the CSE, because of the nonzero net baryon chemical potential,
the strong magnetic field will induce an axial current and a local
axial chemical potential $\mu_{A}$. For example, assuming the reaction plane is on the $y-z$ plane in Fig.\ref{fig:cartoon}, in the $x>0$ or
$x<0$ region, the CSE will lead $\mu_{A}>0$ or $\mu_{A}<0$. When
there exists a local axial chemical potential, the CME will give rise
to the charge separation, where the positive-charged particles will be pushed away from the reaction plane as illustrated in the right panel of Fig.\ref{fig:CSE-and-CME}.
These dynamical and reaction-plane-dependent fluctuations
of electric charge is expected not to vanish when averaged over lots
of events. A possible result from these effects is the charge asymmetry encoded by the $v_{2}$ difference
of $\pi^{\pm}$\cite{Burnier:2011bf}.  

In \cite{Huang:2013iia}, the authors considered a
small global axial chemical potential induced by fluctuations or topological
transitions of QCD vacuum in each events. For example, as shown in
Fig.\ref{fig:CESE-and-CME}, we assume there is a global $\mu_{A}>0$
in a certain event. In the Cu+Au collisions, because of geometric asymmetry
of the nuclei, there will be a large electric field from Au to Cu in the
early stage\cite{Hirono:2012rt}, e.g. as
shown in Fig.\ref{fig:CESE-and-CME}, the $\mathbf{E}$ field is
along the $y$ direction. Because of the normal electric conduction $\mathbf{j}_{v}\propto\mathbf{E}$
, the positive and negative charged particles will be dragged to the $y>0$
and $y<0$ region, respectively. However, since the CESE yields $\mathbf{j}_{a}\propto\mu_{V}\mu_{A}\mathbf{E}$,
the right and left handed quarks will also be pushed to the $y>0$
and $y<0$ region, respectively. Therefore, the electric field enhances
the charge and chirality separation. 
Now in $y>0$
region, there are more positive-charged particles and more right-handed
particles, i.e. locally $\mu_{V}>0,\mu_{A}>0$. While in $y<0$ region,
there are more negative-charged particles and more left-handed particles,
i.e. locally $\mu_{V}<0,\mu_{A}<0$. 

Now we can add the CME and CSE to the system.
As shown in the right panel of Fig.\ref{fig:CESE-and-CME}, in the $y>0$ region, since
$\mathbf{j}_v\propto\mu_{A}\mathbf{B}$ and $\mathbf{j}_{a}\propto\mu_{V}\mathbf{B}$
with $\mu_{V},\mu_{A}>0$, the positive-charged and right-handed (or negative-charged and left-
handed) quarks will move
along (or along the opposite direction of) the $\mathbf{B}$ field
and accumulate in the $x>0$ (or $x<0$) side. Similarly,
in the $y<0$ region, the opposite processes will occur because of $\mu_{V},\mu_{A}<0$. 
In $x>0$ sider
of $y<0$ region, the positive-charged and right-handed particles
will move along the opposite direction of the $\mathbf{B}$ field.
Note that, initially there is the net $\mu_{V}>0$ after the collisions.
Therefore, after the evolution in $x>0$ side there will still be
more positive-charged particles at $y>0$ region than negative-charged
particles at $y<0$ region.
Eventually, the combinations of magnetic and electric fields might
cause a quadrupole distribution at certain angle $\Psi_{q}$ with respect to the reaction plane. 

The Hall and chiral
Hall effects are expected to play a role in such strong electric and
magnetic fields. However, the dynamics evolution is very complicated
and the quantitative predictions require numerical
studies in hydrodynamics. Here, we will only discuss some possible
phenomena in a qualitative description. For simplicity,
we neglect all other chiral effects expect Hall and chiral Hall effects.
As illustrated in Fig. \ref{fig:Hall-and-chiral}, in heavy ion collisions,
the fireball is approximately boost-invariant along the $z$ direction as the beam direction in Fig.\ref{fig:Hall-and-chiral}. Since both magnetic and electric
fields are at the transverse plane ($x,y$) plane in Fig. \ref{fig:Hall-and-chiral},
according to (\ref{eq:Hall_strong_B_01}) and
(\ref{eq:power_counting_hall_strong_02}), the
Hall and chiral Hall effects will only induce currents anti-parallel
or parallel to the $z$ direction. For example, we
assume there is a global net $\mu_{A}>0$ and $\mu_{V}>0$ in the
QGP. Since $j_{v,z}\propto-n_{v}\propto-\mu_{V}$, the positive-charged
particles will move anti-parallel to $z$ direction, while the negative-charged
particles will move parallel to $z$ direction. From $j_{a,z}\propto-n_{a}\propto-\mu_{A}$,
the chirality separation happens similarly. It will further causes the nontrivial charge distribution with rapidity. Note that an axial Hall current can be generated by the CHE even at $\mu_V=0$. 
Furthermore, when combining the CESE, CME, and CHE, we might find the difference in charge asymmetry of the flow coefficients $v_n$ of charged pions with different rapidity. For example, we could expect that the quadrapole distribution will be enhanced in the backward rapidity but reduced in the forward rapidity.


In the next section, we will study the propagating waves coming from the density fluctuations and the above effects, while we only consider the fluctuations of currents
and then solve the linearized desperation relation and discuss all possible
propagating modes. We will leave the numerical studies based on hydrodynamic simulations in the future. 

\section{Density Waves with finite chemical potentials}\label{density_waves}
\subsection{Chiral Magnetic Waves}
We firstly review the derivation of CMW from the CME and CSE in the right-handed and left-handed (R/L) bases in the presence of an external magnetic field. However, we will consider the presence of nonzero chemical potentials and electric conductivities of the medium. The CME and CSE along with the internal electric fields yield
\begin{eqnarray}\label{jRL}
{\bf j_R}=\lambda\mu_R{\bf B}+e\sigma_R{\bf E}_{in},\quad{\bf j_L}=-\lambda\mu_L{\bf B}+e\sigma_L{\bf E}_{in},
\end{eqnarray}
where $\lambda=N_ce/(2\pi^2)$ and $\sigma_{R/L}$ denote the electric conductivities for right/left handed fermions and ${\bf B}$ denotes a constant strong background magnetic field. Therefore, the fluctuations of magnetic fields from the charged particles could be neglected. For simplicity, we further consider a decoupled system, where the right-handed fermions do not interact with the left-handed fermions. The ${\bf E}_{in}$ here represents an "internal" electric field, which may come from a charged medium. 
Given that the right-handed fermions do not interact with left-handed fermions, we may assume that $\mu_R(\sigma_R)$ and $\mu_L(\sigma_L)$ depend on $j^0_R$ and $j^0_L$, respectively. By implementing the conservation equation $\partial_{\mu}j^{\mu}=0$ and $\nabla\cdot{\bf B}=0$, $\nabla\cdot{\bf E_{in}}=j^0_v$, we obtain
\begin{eqnarray}\nonumber\label{waveeqRL0}
&&\partial_0 j^0_R+\lambda {\bf B\cdot\nabla}\mu_R+e\sigma_Rj^0_v+e{\bf E_{in}\cdot\nabla}\sigma_R=0,\\
&&\partial_0 j^0_L-\lambda {\bf B\cdot\nabla}\mu_L+e\sigma_Lj^0_v+e{\bf E_{in}\cdot\nabla}\sigma_L=0.
\end{eqnarray} 
We then introduce the fluctuations of the charge densities in R/L bases,
\begin{eqnarray}
j^0_{R/L}\rightarrow n_{R/L}+\delta j_{R/L}^0.
\end{eqnarray}

Inserting the static charge densities $n_{R/L}$ or $n_{v/a}$ back
to (\ref{waveeqRL0}) and assuming $\sigma_{R/L}$ and $\mu_{R/L}$
uniform, we can solve the charge densities directly, i.e. $n_{v}=n_{0,v}\exp\left(-e\sigma_{v}t\right)+const.$,
with $n_{0,v}$ constant given by initial conditions. That implies
the nonzero charge density will eventually damp out with the damping
time $\tau_{c}=1/(e\sigma_{v})$, which was as well indicated
in \cite{Huang:2013iia}. Therefore, the time scale of the fluctuations
$\delta j_{R/L}$ or $\delta j_{v/a}$ is required to be much smaller
than the damping time $\tau_{c}$. Fortunately, we find in
the following model used in Sec. \ref{CHE_holography}, the damping
time scale is about a few $fm/c$. 

By using the results in our previous study of the DC conductivities
in holography in \cite{Pu:2014cwa}, we get $e\sigma_{v}\sim5T\hat{\sigma}_{v}$
with $\hat{\sigma}_{v}$ being a dimensionless constant depending
on the ratios of vector and axial chemical potentials to temperature.
When $T=200$ MeV as the average temperature in RHIC, we obtained
$e\sigma_{v}\sim26$ MeV for $\mu_{V}=\mu_{A}=0$ and $e\sigma_{v}\sim36$
MeV for $\mu_{V}=4T$ and $\mu_{A}=0$. The corresponding characteristic
times are $\tau_{c}\sim7.6fm/c$ and $\tau_{c}\sim5.5fm/c$, respectively.
These values of the damping times are sufficient long to compare
with the fluctuations we assumed here. In this case, we can just simply
consider $n_{R/L}$ or $n_{v/a}$ as constants in our following discussion.
Similarly, according to the lattice calculations\cite{Aarts:2007wj,Ding:2010ga,Tuchin:2013ie},
the DC conductivity of a static QGP is $e\sigma_{v}\sim5.8T/T_{c}$
MeV with $T_{c}$ the critical temperature. The damping time scale
is about $\tau_{c}=1/(e\sigma_{v})\sim17-34fm/c$ for $T\sim T_{c}-2T_{c}$
as the temperature of the QGP in RHIC.


From (\ref{jRL}), we find
\begin{eqnarray}
&&{\bf\delta j_R}=\lambda\alpha_R\delta j^0_R{\bf B}+e\beta_{R}\delta j^0_{R}{\bf E_{in}},\quad{\bf\delta j_L}=-\lambda\alpha_L\delta j^0_L{\bf B}+e\beta_{L}\delta j^0_{L}{\bf E_{in}},
\end{eqnarray}
where
\begin{eqnarray}
\alpha_{R/L}=\left(\frac{\partial\mu_{R/L}}{\partial j^0_{R/L}}\right)_{j^0_{R/L}\rightarrow n_{R/L}},\quad \beta_{R/L}=\left(\frac{\partial\sigma_{R/L}}{\partial j^0_{R/L}}\right)_{j^0_{R/L}\rightarrow n_{R/L}}.
\end{eqnarray}
By assuming an uniform charge distribution, where $n_{R/L}$ are spacetime independent, (\ref{waveeqRL0}) becomes
\begin{eqnarray}\nonumber\label{waveeqRL}
&&\partial_0\delta j^0_R+\lambda \alpha_R{\bf B\cdot\nabla}\delta j^0_R
+e\beta_Rn_v\delta j^0_R+e\sigma_R\delta j^0_R+e\beta_R{\bf E_{in}\cdot\nabla}\delta j^0_R=0,\\
&&\partial_0\delta j^0_L-\lambda \alpha_L{\bf B\cdot\nabla}\delta j^0_L
+e\beta_Ln_v\delta j^0_L+e\sigma_L\delta j^0_L+e\beta_L{\bf E_{in}\cdot\nabla}\delta j^0_L=0.
\end{eqnarray}
Here we assume that $\mu_{R/L}$ and $\sigma_{R/L}$ have no spacial dependence, while their fluctuations do.
For ${\bf E_{in}\ll{\bf B}}$, we may drop the last terms explicitly depending on the electric field, whereas we could preserve the terms contributed by nonzero $\beta_{R/L}$ and $\sigma_{R/L}$.  
We may now rewrite (\ref{waveeqRL}) in terms of the vector/axial(v/a) bases, which reads
\begin{eqnarray}\nonumber\label{eq:perturbation_01}
&&\partial_0\delta j^0_v+\lambda (\alpha_{-}{\bf B\cdot\nabla}\delta j^0_v+\alpha_{+}{\bf B\cdot\nabla}\delta j^0_a)
+en_v(\beta_+\delta j_v^0+\beta_-\delta j_a^0)+e\sigma_v\delta j^0_v=0,\\
&&\partial_0\delta j^0_a+\lambda (\alpha_{-}{\bf B\cdot\nabla}\delta j^0_a+\alpha_{+}{\bf B\cdot\nabla}\delta j^0_v)
+en_v(\beta_-\delta j_v^0+\beta_+\delta j_a^0)+e\sigma_a\delta j^0_v=0,
\end{eqnarray}
where
\begin{eqnarray}
\delta j^{\mu}_{v/a}=\frac{1}{2}(\delta j^{\mu}_R\pm \delta j^{\mu}_L),\quad \alpha_{\pm}=\frac{1}{2}(\alpha_R\pm\alpha_L),\quad
\beta_{\pm}=\frac{1}{2}(\beta_R\pm\beta_L),\quad 
\sigma_{v/a}=\frac{1}{2}(\sigma_R\pm\sigma_L).
\end{eqnarray}
By taking $\delta j^0_{v/a}=C_{v/a}e^{-iwt+i{\bf k\cdot x}}$ with $C_{v/a}$ being constants, we derive the dispersion relation 
\begin{eqnarray}\label{CMWomega}
\omega_{\pm}=\lambda\alpha_-{\bf B\cdot k}-ien_v\beta_+-\frac{ie\sigma_v}{2}\pm 
\sqrt{\left(\lambda\alpha_+{\bf B\cdot k}-ien_v\beta_-\right)\left(\lambda\alpha_+{\bf B\cdot k}-ie(n_v\beta_-+\sigma_a)\right)-\frac{e^2\sigma_v^2}{4}},
\end{eqnarray}
where $C_a=\pm C_v$.
In the hydrodynamic description, we may make a small-momentum expansion of the right hand side in (\ref{CMWomega}),
\begin{eqnarray}\nonumber\label{CMWdispersion}
\omega_{\pm}&=& -ie\left(n_v \beta_{+}+\frac{\sigma_v}{2}\right)\mp ie\sqrt{n_v^2 \beta_-^2+ n_v\beta_-\sigma_a+\frac{\sigma_v^2}{4}}+ \lambda  \left(\alpha_-\pm\frac{\alpha_+(2n_v\beta_{-}+\sigma_a)}{\sqrt{4n_v^2\beta_-^2+4 n_v \beta_- \sigma_a+\sigma_v^2}}\right){\bf B\cdot k}
\\
&&\pm\frac{i \alpha_+^2 \lambda^2 \left(\sigma_v^2-\sigma_a^2\right) ({\bf B\cdot k})^2}{e\left(4 n_v^2\beta_-^2+4 n_v \beta_- \sigma_a+\sigma_v^2\right)^{3/2}}
+\mathcal{O}\left({({\bf B\cdot k})^3}\right).
\end{eqnarray}
The momentum-independent terms above characterize the damping effect and the prefactors of the terms linear to ${\bf k}$ corresponds to the wave velocity. 
The last term proportional to ${\bf k}^2$ is associated with the diffusion.
   
For a chargeless system ($n_v=0$), the two modes become
\begin{eqnarray}\nonumber
\omega_+&=&-ie\sigma_v+\lambda  \left(\alpha_{-} +\alpha_{+}\frac{\sigma_a}{\sigma_v}\right){\bf B\cdot k}+i(e\sigma_v)^{-1}\alpha_+^2 \lambda ^2 \left(1-\frac{\sigma_a^2}{\sigma_v^2}\right) ({\bf B\cdot k})^2+\mathcal{O}\left({({\bf B\cdot k})^3}\right),\\
\omega_-&=& \lambda  \left(\alpha_{-}-\alpha_{+}\frac{\sigma_a}{\sigma_v}\right) {\bf B\cdot k}-i(e\sigma_v)^{-1}\alpha_+^2 \lambda ^2 \left(1-\frac{\sigma_a^2}{\sigma_v^2}\right) ({\bf B\cdot k})^2+\mathcal{O}\left({({\bf B\cdot k})^3}\right).
\end{eqnarray}

In the limit of $n_v=0$ and $\sigma_{v/a}=0$, the dispersion relation in (\ref{CMWomega}) further reduces to
\begin{eqnarray}\label{dispCMW1}
\omega_{\pm}=\lambda({\bf B\cdot k})(\alpha_-\mp\alpha_+)=-\lambda({\bf B\cdot k})\alpha_L\mbox{ or }
\lambda({\bf B\cdot k})\alpha_R.
\end{eqnarray}
It turns out that there exist two wave velocities $v_{\chi}=N_c|{\bf eB}|\alpha_{R/L}/(2\pi^2)$. For small chemical potentials (small charge densities), 
$\alpha_R=\alpha_L$, the two velocities become degenerate. Our result then reduces to what has been found in \cite{Kharzeev:2010gd}.

\subsection{Chiral Electric Waves}
Generally, in a QCD plasma, the interaction between left and right
handed fermions will play a role to the propagating modes. However,
since we will only investigate those modes by SS model, in which there
are no effective interactions between the fermions with different
chiralities, we will neglect this kind of interaction in
the following discussion, i.e. we assume $\sigma_{R}$ (or $\sigma_{L}$)
will only be functions of $T$ and $\mu_{R}$ (or $\mu_{L}$), respectively.
 
By following the same strategy, we can derive the CEW in the presence of an external electric field. We may start with
\begin{eqnarray}
{\bf j_R}=e\sigma_R(\mu_R){\bf E}=e\sigma_R(j^0_R){\bf E},\quad{\bf j_L}=e\sigma_L(\mu_L){\bf E}=e\sigma_L(j^0_L){\bf E}.
\end{eqnarray}
In general, we set $\bf E=E_{ex}+E_{in}$, where $\bf{E_{ex}}$ and $\bf{E_{in}}$ denote the external and internal electric fields, respectively. We may assume that the external electric field is a constant field, whereas ${\bf \nabla\cdot E_{in}}=j^0_v$.
Similarly, we introduce the fluctuations of the currents,
\begin{eqnarray}\label{flucj}
{\bf\delta j_{R/L}}=e\beta_{R/L}\delta j^0_{R/L}{\bf E}.
\end{eqnarray}
The conservation equation $\partial_{\mu}j^{\mu}=0$ then leads to
\begin{eqnarray}
\partial_0j^0_{R/L}+e{\bf E\cdot\nabla}\sigma_{R/L}+e\sigma_{R/L}{\bf\nabla\cdot E}=0.
\end{eqnarray}
By further perturbing the above equation and utilizing $\nabla\cdot{\bf E}=j_v^0$ and $\delta\sigma_{R/L}=\beta_{R/L}\delta j^0_{R/L}$, we find
\begin{eqnarray}\label{waveRL}
\partial_0\delta j^0_{R/L}+e\beta_{R/L}{\bf E\cdot\nabla}\delta j^0_{R/L}+e\beta_{R/L}n_v\delta j^0_{R/L}+e\sigma_{R/L}\delta j^0_v=0.
\end{eqnarray}
Here $\mathbf{E}$ in the above equation is the total electric field.
In a strong external field case, the contribution from $\mathbf{E_{ex}}$
is dominant, where the one from $\mathbf{E_{in}}$
can be neglected. However, in the absence of external fields, $\mathbf{E_{in}}$
becomes dominant. Actually, in this
case, this term plays an important role to guarantee the conservation
of the total charge number. Especially, in the $n_{v}=0$ limit, this
term will be proportional to $\mathbf{E_{in}}\cdot\mathbf{k}$ and
finally appear in (34). Although it will be subleading in terms of the fluctuations in the bulk, it will be in the order linear to $\delta j^0_{R/L}$ on the surface of the medium, which yields the propagation of density waves outward the thermal medium. The argument for the Hall current in
(40) is similar.  

We may further rewrite (\ref{waveRL}) in terms of the v/a bases,
\begin{eqnarray}\nonumber\label{waveva}
&&\partial_0\delta j^0_v+e (\beta_{+}{\bf E\cdot\nabla}\delta j^0_v+\beta_{-}{\bf E\cdot\nabla}\delta j^0_a+n_v\beta_+\delta j^0_v+n_v\beta_-\delta j^0_a+\sigma_v\delta j^0_v)=0,\\
&&\partial_0\delta j^0_a+e (\beta_{-}{\bf E\cdot\nabla}\delta j^0_v+\beta_{+}{\bf E\cdot\nabla}\delta j^0_a+n_v\beta_-\delta j^0_v+n_v\beta_+\delta j^0_a+\sigma_a\delta j^0_v)=0.
\end{eqnarray}
When taking $\delta j^0_{v/a}=C_{v/a}e^{-iwt+i{\bf k\cdot x}}$ with $C_{v/a}$ being constants, the dispersion relation reads 
\begin{eqnarray}\label{CEWomega}
\omega_{\pm}=e\beta_+{\bf E\cdot k}-ien_v\beta_+-\frac{ie\sigma_v}{2}\pm e
\sqrt{\left(\beta_-{\bf E\cdot k}-in_v\beta_-\right)\left(\beta_-{\bf E\cdot k}-i(n_v\beta_-+\sigma_a)\right)-\frac{\sigma_v^2}{4}}.
\end{eqnarray}
By expanding (\ref{CEWomega}) with the momentum in the hydrodynamic approximation, we obtain
\begin{eqnarray}\nonumber
\omega_{\pm}&=& -ie \left(n_v \beta_{+}+\frac{\sigma_v}{2}\pm\sqrt{n_v^2\beta_-^2+n_v \beta_-\sigma_a+\frac{\sigma_v^2}{4}}\right)+
e \left(\beta_{+}\pm\frac{ \beta_- (2 n_v \beta_{-}+\sigma_a)}{\sqrt{4n_v^2 \beta_-^2+4 n_v\beta_{-}\sigma_a+\sigma_v^2}}\right){\bf E\cdot k}\\
&&\pm\frac{ie \beta_-^2 \left(\sigma_v^2-\sigma_a^2\right) ({\bf E\cdot k})^2}{\left(4n_v^2 \beta_-^2+4n_v\beta_-\sigma_a+\sigma_v^2\right)^{3/2}}
+\mathcal{O}\left({({\bf E\cdot k})^3}\right).
\end{eqnarray} 
Similar to CMW, for a chargeless system ($n_v=0$), we find two modes,
\begin{eqnarray}\nonumber
\omega_+&=&-ie\sigma_v+e\left(\beta_{+} +\beta_{-}\frac{\sigma_a}{\sigma_v}\right){\bf E\cdot k}+ie\sigma_v^{-1}\beta_-^2 \left(1-\frac{\sigma_a^2}{\sigma_v^2}\right) ({\bf E\cdot k})^2+\mathcal{O}\left({({\bf E\cdot k})^3}\right),\\
\omega_-&=&e\left(\beta_{+}-\beta_{-}\frac{\sigma_a}{\sigma_v}\right) {\bf E\cdot k}-ie\sigma_v^{-1}\beta_-^2 \left(1-\frac{\sigma_a^2}{\sigma_v^2}\right) ({\bf E\cdot k})^2+\mathcal{O}\left({({\bf E\cdot k})^3}\right).
\end{eqnarray}
When considering the chargeless case ($n_v=0$, $\sigma_{v/a}=0$), the dispersion relation in (\ref{CEWomega}) reduces to
\begin{eqnarray}\label{dispCEW1}
\omega_{\pm}=e({\bf E\cdot k})(\beta_+\mp\beta_-)=-e({\bf E\cdot k})\beta_L\mbox{ or }
e({\bf E\cdot k})\beta_R.
\end{eqnarray}
This result is very similar to that for CMW. Although the wave velocity of CEW is dictated by the fluctuations of the conductivities, it implicitly depends on the fluctuations of the chemical potentials which influence the conductivities.  

We may now consider the CEW in the limit of small chemical potentials. In light of the assumption in \cite{Huang:2013iia} based on the symmetries, the currents in R/L bases are
\begin{eqnarray}\label{JRLsmallmu}
{\bf j_{R/L}}=e(\sigma_0+\rho\mu_{R/L}^2){\bf E},
\end{eqnarray}
where $\rho$ is a function of temperature. Note that we drop the interaction between the R/L sectors, which is interpreted as the screening in \cite{Huang:2013iia}. From (\ref{JRLsmallmu}), the CESE is given by
\begin{eqnarray}\nonumber
{\bf j_v}&=&e\left(\sigma_0+\rho(\mu_v^2+\mu_a^2)\right){\bf E},\\
{\bf j_a}&=&e\chi_e\mu_v\mu_a{\bf E},
\end{eqnarray}
where $\chi_e=2\rho$. Given that $\mu_{R/L}=\alpha_{R/L}j^0_{R/L}$
\footnote{Although $\delta\mu_{R/L}=\alpha_{R/L}\delta j^0_{R/L}$ is always true, $\mu_{R/L}=\alpha_{R/L}j^0_{R/L}$ only hold for $n_{R/L}$ being small or $n_{R/L}$ being linearly dependent to $\mu_{R/L}$.}, which corresponds to the case with small densities, we obtain
\begin{eqnarray}
\beta_{R/L}=2\rho\alpha_{R/L}^2n_{R/L}.
\end{eqnarray}
For small chemical potentials, we have $\alpha_R=\alpha_L=\alpha_+$, which yields
\begin{eqnarray}\label{betapm}
\beta_+=2\rho\alpha_+^2n_v,\quad\beta_-=2\rho\alpha_+^2n_a.
\end{eqnarray}
The wave equations in (\ref{waveva}) up to $\mathcal{O}(n_{v/a})$ now reduces to
\begin{eqnarray}\nonumber\label{wavevasmallmu}
&&\partial_0\delta j^0_v+2e\rho\alpha_+^2 (n_v{\bf E\cdot\nabla}\delta j^0_v+n_a{\bf E\cdot\nabla}\delta j^0_a)+e\sigma_0\delta j^0_v=0,\\
&&\partial_0\delta j^0_a+2e\rho\alpha_+^2 (n_a{\bf E\cdot\nabla}\delta j^0_v+n_v{\bf E\cdot\nabla}\delta j^0_a)=0.
\end{eqnarray}
We may compare (\ref{wavevasmallmu}) with the result found in \cite{Huang:2013iia}. By definitions, we find
\begin{eqnarray}\nonumber
\alpha_v&=&\frac{\partial\mu_v}{\partial j^0_v}=\frac{1}{2}\left(\frac{\partial\mu_R}{\partial j^0_R}\frac{\partial j^0_R}{\partial j^0_v}+\frac{\partial\mu_L}{\partial j^0_L}\frac{\partial j^0_L}{\partial j^0_v}\right)=\alpha_+,\\
\alpha_a&=&\frac{\partial\mu_a}{\partial j^0_a}=\frac{1}{2}\left(\frac{\partial\mu_R}{\partial j^0_R}\frac{\partial j^0_R}{\partial j^0_a}-\frac{\partial\mu_L}{\partial j^0_L}\frac{\partial j^0_L}{\partial j^0_a}\right)=\alpha_+=\alpha_v.
\end{eqnarray}
When turning off the magnetic field and taking $\chi_e=2\rho$ and $\rho=\sigma_2$ as defined in \cite{Huang:2013iia}, 
we find that (\ref{wavevasmallmu}) is consistent with the result therein in the absence of a magnetic field.

By further including the Hall effect yet excluding CME and CSE, the fluctuations of the currents become
\begin{eqnarray}
(\delta j_{R/L})_i=e(\beta_{R/L})_{ij}\delta j^0_{R/L}E_j,
\end{eqnarray}
where
\begin{eqnarray}
(\beta_{R/L})_{ij}=\left(\frac{\partial(\sigma_{R/L})_{ij}}{\partial j^0_{R/L}}\right)_{j^0_{R/L}\rightarrow n_{R/L}}.
\end{eqnarray}
The wave equations now take the form
\begin{eqnarray}
\partial_0\delta j^0_{R/L}+e(\beta_{R/L})_{ij}E_j\partial_i\delta j^0_{R/L}+e(\beta_{R/L})_{ii}n_v\delta j^0_{R/L}+e(\sigma_{R/L})_{ii}\delta j^0_v=0.
\end{eqnarray}
We can subsequently work in the $v/a$ bases and derive the dispersion relations. 
By taking $\delta j^0_{v/a}=C_{v/a}e^{-iwt+i{\bf k\cdot x}}$ with $C_{v/a}$ being constants, the dispersion relation reads 
\begin{eqnarray}\label{CEWomegaij}
\omega_{\pm}&=&e(\beta_+)_{ij}E_jk_i-ien_v(\beta_+)_{ii}-\frac{ie(\sigma_v)_{ii}}{2}
\\\nonumber
&&\pm e
\sqrt{\left((\beta_-)_{ij} E_jk_i-in_v(\beta_-)_{ii}\right)\left((\beta_-)_{ij}E_j k_i-i(n_v(\beta_-)_{ii}+(\sigma_a)_{ii})\right)-\frac{e^2(\sigma_v)_{ii}^2}{4}}.
\end{eqnarray}
In our setup, we have
\begin{eqnarray}
(\sigma_{v/a})_{ii}=(\sigma_{v/a})_{yy},\quad (\beta_{\pm})_{ii}=(\beta_{\pm})_{yy},\quad 
(\beta_{\pm})_{ij}E_jk_i=(\beta_{\pm})_{zy}E_yk_z+(\beta_{\pm})_{yy}E_yk_y.
\end{eqnarray}    
After making the momentum-expansion, the dispersion relation in (\ref{CEWomegaij}) becomes
\begin{eqnarray}
\omega_{\pm}=-i\tau_{\pm}^{-1}+(v_{\pm})_ik^i-i(D_{\pm})_{ij}k^ik^j,
\end{eqnarray}
where
\begin{eqnarray}\nonumber
\tau_{\pm}^{-1}&=&e\left(n_v(\beta_+)_{ii}+\frac{(\sigma_v)_{ii}}{2}
\pm\sqrt{n_v^2(\beta_-)_{ii}^2+n_v(\beta_-)_{ii}(\sigma_a)_{ii}+\frac{(\sigma_v)_{ii}^2}{4}}\right),
\\\nonumber
(v_{\pm})_k&=&e\left((\beta_{+})_{kj}\pm\frac{ (\beta_-)_{kj} (2 n_v (\beta_{-})_{ii}+(\sigma_a)_{ii})}{\sqrt{4n_v^2 (\beta_-)_{ii}^2+4 n_v(\beta_{-})_{ii}(\sigma_a)_{ii}+(\sigma_v)_{ii}^2}}\right)E_j,
\\
(D_{\pm})_{ij}&=&\mp\frac{e (\beta_-)_{ik}(\beta_-)_{jl} \left((\sigma_v)_{mm}^2-(\sigma_a)_{mm}^2\right) (E_k E_l)}{\left(4n_v^2 (\beta_-)_{mm}^2+4n_v(\beta_-)_{mm}(\sigma_a)_{mm}+(\sigma_v)_{mm}^2\right)^{3/2}}.
\end{eqnarray}
Here $\tau_{\pm}$ represent the damping times for two modes of the density wave and $(v_{\pm})_k$ correspond to the wave velocities. The diffusion of the density wave is characterized by $(D_{\pm})_{ij}$.
In the following sections, we will employ the SS model in holography to investigate the CESE, CHE, and CEW in the strongly coupled QGP.

\section{SS model}\label{SS_model}
We will follow the approach in \cite{O'Bannon:2007in,Lifschytz:2009si} to investigate the currents induced by the external electromagnetic fields at finite chemical potentials.  
In the SS model at finite temperature, the induced metric of $D8/\overline{D8}$ branes in the chiral symmetric phase is given by
\begin{eqnarray}
ds^2=\left(\frac{U}{L}\right)^{3/2}(-f(U)dt^2+d\vec{x}^2)
+\left(\frac{L}{U}\right)^{3/2}\frac{dU^2}{f(U)}
+\left(\frac{L}{U}\right)^{3/2}U^2d\Omega^2_4,
\end{eqnarray}
where $f(U)=1-U_T^3/U^3$ with $U_T$ being the position of an event horizon and $L=(\pi^3g_sN_cl_s^3)^{1/3}$ is the curvature radius. The temperature of the background reads 
\begin{eqnarray}
T=\frac{3}{4\pi}\left(\frac{U_T}{L^3}\right)^{1/2}.
\end{eqnarray}
There are also background dilaton and form flux
\begin{eqnarray}
e^{\phi}=g_s\left(\frac{U}{L}\right)^{3/4},\quad F_4=\frac{2\pi N_c}{V_4}\epsilon_4,
\end{eqnarray}
where $V_4$ is the volume of the four-sphere and $\epsilon_4$ is the corresponding volume form.
The full DBI action reads
\begin{eqnarray}
S_{DBI}=S_{D8}+S_{\overline{D8}},
\end{eqnarray}
where
\begin{eqnarray}
S_{D8/\overline{D8}}=-T_{D8}\int d^9xe^{-\phi}\sqrt{-\mbox{det}(g+2\pi\alpha'F_{L/R})}.
\end{eqnarray}
Moreover, we have Chern-Simons (CS) terms
\begin{eqnarray}
S_{D8/\overline{D8}}^{CS}=\mp\frac{N_c}{96\pi^2}\int d^4xdU\epsilon^{MNPQR}(A_{L/R})_M(F_{L/R})_{NP}(F_{L/R})_{QR}.
\end{eqnarray}
By turning on the world-volume gauge fields \footnote{Here $E_y$ and $B_x$ actually correspond to $eE_y$ and $eB_x$. We will hereafter omit $e$ in the holographic computations for simplicity.}, $(A_{L/R})_t(U)$, $(A_{L/R})_x(t,U)=(a_{L/R})_x(U)$, $(A_{L/R})_y(t,U)=-E_yt+(a_{L/R})_y(U)$
, and $(A_{L/R})_z(t,U)=B_xy+(a_{L/R})_z(U)$, we obtain
\begin{eqnarray}\nonumber
S_{D8/\overline{D8}}&=&-C\int d^4xdUU^{5/2}\sqrt{X},
\end{eqnarray}
where
\begin{eqnarray}\nonumber
X&=&1+\frac{B_x^2 L^3}{U^3}-\frac{E_y^2 L^3}{U^3 f}-A_t'^2\left(1+\frac{B_x^2 L^3 }{U^3}\right)+a_x'^2f\left(1-\frac{E_y^2 L^3 a_x'^2}{fU^3}+\frac{B_x^2 L^3 }{U^3}\right)+f a_y'^2\\\nonumber
&&+\frac{2 B_x E_y L^3 A_t' a_z'}{U^3}+a_z'^2f\left(1-\frac{E_y^2 L^3}{fU^3}\right),\\
C&=&\frac{T_{D8}V_4L^{3/2}}{g_s}=\frac{N_c}{96\pi^5l_s^6L^{3/2}}.
\end{eqnarray}
Here the primes denote the derivatives with respect to $U$. We also set $2\pi l_s^2=1$ GeV$^{-2}$ and drop the $L/R$ subscripts above for simplicity. In our setup, the CS terms read
\begin{eqnarray}
S_{D8/\overline{D8}}^{CS}=\mp\frac{8N_c}{96\pi^2}\int d^4xdU\left(B_x(A_ta_x'-a_xA_t')+E_y(a_xa_z'-a_za_x')\right).
\end{eqnarray}
The full actions take the form
\begin{eqnarray}
S_{D8/\overline{D8}}^f=-C\left(\int d^4xdUU^{5/2}\sqrt{X}\pm r\int d^4xdU\left(B_x(A_ta_x'-a_xA_t')+E_y(a_xa_z'-a_za_x')\right) \right),
\end{eqnarray}
where $r=N_c/(12\pi^2C)=(2\pi l_s)^3L^{3/2}$. We may add the boundary terms according to \cite{Lifschytz:2009si}, which lead to $r=3/2\times (2\pi l_s)^3L^{3/2}$. The value of $r$ actually depends on the renormalization scheme.
The equations of motion are
\begin{eqnarray}\label{conseq}\nonumber
\frac{U^{5/2}\left((A_{L/R})'_t\left(1+\frac{B_x^2L^3}{U^3}\right)-(a_{L/R})'_z\frac{B_xE_yL^3}{U^3}\right)}
{\sqrt{X_{L/R}}}&=&(J_{L/R})_t\mp 2rB_x(a_{L/R})_x
\\\nonumber
\frac{U^{5/2}f(a_{L/R})_x'\left(1-\frac{E_y^2 L^3}{fU^3}+\frac{B_x^2 L^3}{U^3}\right)}{\sqrt{X_{L/R}}}&=&(J_{L/R})_x\mp 2r (B_x (A_{L/R})_t-E_y (a_{L/R})_z)\\\nonumber
\frac{U^{5/2}f(a_{L/R})_y'}{\sqrt{X_{L/R}}}&=&(J_{L/R})_y,\\
\frac{U^{5/2}\left((A_{L/R})'_t\frac{B_xE_yL^3}{U^3}+
f(a_{L/R})'_z\left(1-\frac{E_y^2L^3}{fU^3}\right)\right)}
{\sqrt{X_{L/R}}}&=&(J_{L/R})_z\mp 2rE_y(a_{L/R})_x,
\end{eqnarray}
where $(J_{L/R})_{\mu}$ are integration constants. In the AdS/CFT correspondence, the electromagnetic currents correspond to the boundary currents of the DBI actions. From the definition of boundary currents,
\begin{eqnarray}
j_{\mu}=J^b_{\mu}=\frac{\delta S_{EOM}}{\delta A_{\mu}(\infty)}=\left(\frac{\delta L_{eff}}{\delta A'_{\mu}}\right)_{U\rightarrow\infty},
\end{eqnarray}
we have 
\begin{eqnarray}\label{bcurrent}\nonumber
(J^b_{L/R})_t&=&C\left(\frac{U^{5/2}\left((A_{L/R})'_t\left(1+\frac{B_x^2L^3}{U^3}\right)-(a_{L/R})'_z\frac{B_xE_yL^3}{U^3}\right)}
{\sqrt{X_{L/R}}}\pm rB_x(a_{L/R})_x\right)_{U\rightarrow\infty},
\\\nonumber
(J^b_{L/R})_x&=&C\left(-\frac{U^{5/2}f(a_{L/R})_x'\left(1-\frac{E_y^2 L^3}{fU^3}+\frac{B_x^2 L^3}{U^3}\right)}{\sqrt{X_{L/R}}}\mp r (B_x (A_{L/R})_t-E_y (a_{L/R})_z)\right)_{U\rightarrow\infty},
\\\nonumber
(J^b_{L/R})_y&=&C\left(-\frac{U^{5/2}f(a_{L/R})_y'}{\sqrt{X_{L/R}}}\right)_{U\rightarrow\infty},
\\
(J^b_{L/R})_z&=&C\left(-\frac{U^{5/2}\left((A_{L/R})'_t\frac{B_xE_yL^3}{U^3}+
f(a_{L/R})'_z\left(1-\frac{E_y^2L^3}{fU^3}\right)\right)}
{\sqrt{X_{L/R}}}\mp rE_y(a_{L/R})_x\right)_{U\rightarrow\infty},
\end{eqnarray}
where $L_{eff}$ is the effective Lagrangian.
By comparing (\ref{conseq}) and (\ref{bcurrent}), the boundary currents can be rewritten as
\begin{eqnarray}\nonumber
(J^b_{L/R})_t&=&C\left((J_{L/R})_t\mp r B_x(a_{L/R})_x\right)_{U\rightarrow\infty},
\\\nonumber
(J^b_{L/R})_x&=&C\left(-(J_{L/R})_x\pm r \left(B_x(A_{L/R})_t-E_y(a_{L/R})_z\right)\right)_{U\rightarrow\infty},
\\\nonumber
(J^b_{L/R})_y&=&-C(J_{L/R})_y,
\\
(J^b_{L/R})_z&=&C\left(-(J_{L/R})_z\pm rE_y(a_{L/R})_x\right)_{U\rightarrow\infty}.
\end{eqnarray}
Following \cite{Lifschytz:2009si}, we may define the modified currents,
\begin{eqnarray}\nonumber
(\tilde{J}_{L/R})_t&=&(J_{L/R})_t\mp 2rB_x(a_{L/R})_x,\\\nonumber
(\tilde{J}_{L/R})_x&=&(J_{L/R})_x\mp 2r (B_x (A_{L/R})_t-E_y (a_{L/R})_z),\\\nonumber
(\tilde{J}_{L/R})_y&=&(J_{L/R})_y,\\
(\tilde{J}_{L/R})_z&=&(J_{L/R})_z\mp 2rE_y(a_{L/R})_x.
\end{eqnarray}
By doing some algebra with (\ref{conseq}), we find 
\begin{eqnarray}\label{Ap}\nonumber
(A_{L/R})'_t&=&\pm\frac{\Big|\left(1-\frac{E_y^2L^3}{fU^3}\right)(\tilde{J}_{L/R})_t+\frac{E_yB_xL^3}{fU^3}(\tilde{J}_{L/R})_z\Big|}{\sqrt{Z}},\\
(A_{L/R})'_x&=&\pm\frac{\Big|(\tilde{J}_{L/R})_x\Big|}{f\sqrt{Z}},\\\nonumber
(A_{L/R})'_y&=&\pm\frac{\Big|\left(1+\frac{B_x^2L^3}{U^3}-\frac{E_y^2L^3}{fU^3}\right)(\tilde{J}_{L/R})_y\Big|}{f\sqrt{Z}},\\\nonumber
(A_{L/R})'_z&=&\pm\frac{\Big|\left(1+\frac{B_x^2L^3}{U^3}\right)(\tilde{J}_{L/R})_z-\frac{E_yB_xL^3}{U^3}(\tilde{J}_{L/R})_t\Big|}{f\sqrt{Z}},
\end{eqnarray}
where
\begin{eqnarray}\label{Zfun}\nonumber
Z&=&\left(1+\frac{B_x^2L^3}{U^3}-\frac{E_y^2L^3}{fU^3}\right)
\left(U^5+(\tilde{J}_{L/R})_t^2-\frac{(\tilde{J}_{L/R})_y^2+(\tilde{J}_{L/R})_z^2}{f}\right)
\\&&
-\frac{L^3}{U^3}\left(B_x(\tilde{J}_{L/R})_t-\frac{E_y(\tilde{J}_{L/R})_z}{f}\right)^2
-\frac{(\tilde{J}_{L/R})_x^2}{f}
\end{eqnarray} 
By requiring that $(A_{L/R})'_{\mu}$ are real and well defined, we have to make both the numerators and denominators on the left hand side of (\ref{Ap}) vanish at a critical point $U=U_c$. We thus have
\begin{eqnarray}\label{Jeom}\nonumber
\left(1-\frac{E_y^2L^3}{fU_c^3}\right)(\tilde{J}_{L/R})_t-\frac{E_yB_xL^3}{fU_c^3}(\tilde{J}_{L/R})_z=0,
\\\nonumber
(\tilde{J}_{L/R})_x=0,\\\nonumber
\left(1+\frac{B_x^2L^3}{U_c^3}-\frac{E_y^2L^3}{fU_c^3}\right)=0,\\\nonumber
\left(1+\frac{B_x^2L^3}{U_c^3}\right)(\tilde{J}_{L/R})_z-\frac{E_yB_xL^3}{U_c^3}(\tilde{J}_{L/R})_t=0,\\
Z(U_c)=0.
\end{eqnarray}
Note that the first equation in (\ref{Jeom}) is redundant, which can be obtained from the third and fourth equations therein. 
In fact, (\ref{Jeom}) is equivalent to finding the double zeros of $Z(U_c)$ from the expression in (\ref{Zfun}), where all three terms therein have double zeroes at $U_c$.
From the third equation in (\ref{Jeom}), we find the critical point
\begin{eqnarray}
U_c=\frac{U_T}{2^{1/3}}\Bigg(1+\frac{L^3}{U^3_T}(E_y^2-B_x^2)
+\sqrt{\frac{4B_x^2L^3}{U_T^3}+\left(1+\frac{L^3}{U^3_T}(E_y^2-B_x^2)\right)^2}\Bigg)^{1/3}
\end{eqnarray}
One may now solve rest of equations in (\ref{Jeom}) to derive $(J_{L/R})_i$ for $i=x,y,z$ in terms of $(J_{L/R})_t$. We find
\begin{eqnarray}\label{JRLbases}\nonumber
(J_{L/R})_x&=&\pm 2 r(B_x (A_{L/R})_t-E_y (a_{L/R})_z)_{U=U_c},\\\nonumber
(J_{L/R})_y&=&-\frac{E_yL^{3/2}U_c^{3/2}}{B_x^2L^3+U_c^3}
\left((J_{L/R})_t^2+B_x^2 L^3U^2+U^5\mp 4r B_x (J_{L/R})_t (a_{L/R})_x+4 B_x^2 r^2 (a_{L/R})_x^2\right)^{1/2}_{U=U_c},\\
(J_{L/R})_z&=&\left(\frac{B_x E_y (J_{L/R})_t L^3\pm 2 r E_y U^3  (a_{L/R})_x}{B_x^2 L^3+U^3}\right)_{U=U_c}.
\end{eqnarray}
The boundary currents then become
\begin{eqnarray}\nonumber\label{bcurrentRL}
(J^b_{L/R})_x&=&C\left[\mp 2 r\left(B_x (A_{L/R})_t-E_y (a_{L/R})_z\right)_{U=U_c}\pm r \left(B_x(A_{L/R})_t-E_y(a_{L/R})_z\right)_{U=\infty}\right],
\\\nonumber
(J^b_{L/R})_y&=&C\Bigg[\frac{E_yL^{3/2}U_c^{3/2}}{B_x^2L^3+U_c^3}
\left((J_{L/R})_t^2+B_x^2 L^3U^2+U^5\mp 4r B_x (J_{L/R})_t (a_{L/R})_x+4 B_x^2 r^2 (a_{L/R})_x^2\right)^{1/2}_{U=U_c}\Bigg],
\\
(J^b_{L/R})_z&=&C\left[\left(\frac{-B_x E_y (J_{L/R})_t L^3\mp 2 r E_y U^3  (a_{L/R})_x}{B_x^2 L^3+U^3}\right)_{U=U_c}\pm \left(rE_y(a_{L/R})_x\right)_{U=\infty}\right].
\end{eqnarray}
In the presence of CS terms, we find that $(J_{L/R})_i$ not only depend on $(J_{L/R})_t$ but also depend on $(a_{L/R})_x$ and $(a_{L/R})_z$ at the boundary and $U_c$. It turns out that the gauge invariance of the boundary currents is broken by the CS terms. The nonzero values of $(a_{R/L})_{i}(\infty)$ with $i=x,y,z$ correspond to the pion gradient in the chiral-symmetry-broken phase\cite{Bergman:2008qv}.  
In the chiral-symmetry-restored phase, $(a_{R/L})_{i}(\infty)$ become free parameters, which are set to zero in \cite{Lifschytz:2009si}. For simplicity and preciseness, we focus on the condition that the particle interaction dominates the topological effect. The axial Hall current should exist without the axial anomaly, while it could vary in the presence of the strong axial anomaly and become non-gauge-invariant in the SS model.     

Considering the gauge-invariant currents from interactions, we may turn off $(a_{L/R})_x(U)$ and neglect the effect from the CS terms.  
By rewriting (\ref{JRLbases}) in terms of vector/axial bases, we find
\begin{eqnarray}\label{bdcurrents}
\nonumber
(J^b_{v/a})_y&=&\frac{CE_yL^{3/2}U_c^{3/2}}{2(B_x^2L^3+U_c^3)}
\left(\sqrt{((J_v)_t+(J_a)_t)^2+B_x^2 L^3U_c^2+U_c^5}\pm
\sqrt{((J_v)_t-(J_a)_t)^2+B_x^2 L^3U_c^2+U_c^5}\right),
\\\nonumber
(J^b_{v/a})_z&=&-C\frac{B_x E_y (J_{v/a})_t L^3}{B_x^2 L^3+U_c^3},
\\
(J^b_{v/a})_t&=&C(J_{v/a})_t.
\end{eqnarray}
Now, both $(J^b_{v/a})_y$ and $(J^b_{v/a})_z$ depend on the charge densities $(J^b_{v/a})_t$ on the boundary as functions of the chemical potentials. To find the relations between the charge densities and the chemical potentials, we have to solve the field equation of $(A_{L/R})_t$ in (\ref{Ap}). By utilizing (\ref{JRLbases}), this field equation can be further written as
\begin{eqnarray}\label{solAt}
(A_{L/R})'_t=\frac{\Big|\left(1-\frac{E_y^2L^3U_c^3}{fU^3(B_x^2L^3+U_c^3)}\right)(J_{L/R})_t\Big|}{\sqrt{Z}}.
\end{eqnarray}
We will then render the boundary conditions $(A_{L/R})_t(U_T)=0$ and numerically solve the field equation. The chemical potentials are given by 
\begin{eqnarray}
\mu_{L/R}=(A_{L/R})_t(\infty),
\end{eqnarray}
which are varied by the values of $(J_{L/R})_t$. 

\section{CESE/CHE in holography}\label{CHE_holography}
\subsection{Weak and Strong Electromagnetic Fields}
Although the boundary currents with different chemical potentials can be solved numerically, we may approximate their analytic expressions in the limit of weak electromagnetic fields.
In the presence of weak electromagnetic fields, the induced currents should follow the linear response theory. When taking $E_y\approx 0$ and $B_x\approx 0$, from (\ref{solAt}), the chemical potentials are given by
\begin{eqnarray}
\frac{\mu_{L/R}}{U_T}=\frac{2}{3\tilde{U}_{L/R}^{5/2}}{}_2F_1
\left(\frac{3}{10},\frac{1}{2},\frac{13}{10},-\frac{1}{\tilde{U}_{L/R}^5}\right),
\quad 
\tilde{U}_{L/R}=\frac{U_T}{(J_{L/R})_t^{2/5}}.
\end{eqnarray}
In the limit of $\tilde{U}_{L/R}\rightarrow 0$, which corresponds to high-density or low-temperature conditions, we find
\begin{eqnarray}
\frac{\mu_{L/R}}{U_T}\approx\frac{2\Gamma\left(\frac{1}{5}\right)\Gamma\left(\frac{13}{10}\right)}{3\sqrt{\pi}\tilde{U}_{L/R}}
-\frac{10\Gamma\left(\frac{13}{10}\right)}{3\Gamma\left(\frac{3}{10}\right)}
+\mathcal{O}(\tilde{U}_{L/R}^5).
\end{eqnarray}
Up to the leading order in the expansion with respect to $\tilde{U}_{L/R}$, we obtain
\begin{eqnarray}
(J_{L/R})_t=\left(\frac{3\sqrt{\pi}}{2\Gamma\left(\frac{1}{5}\right)\Gamma\left(\frac{13}{10}\right)}\right)^{5/2}\mu_{L/R}^{5/2}.
\end{eqnarray}
By expanding the boundary currents in (\ref{bdcurrents}), we derive the relation between the currents and chemical potentials in the high-density(low temperature) limit. The currents now take the form
\begin{eqnarray}\nonumber
(J^b_{v/a})_y&=&
\frac{CE_y}{2}\left(\frac{R}{U_T}\right)^{3/2}\left((J_R)_t\pm(J_L)_t\right)
=\frac{CE_y}{2a^3T^3L^3}
\left(\frac{3\sqrt{\pi}}{2\Gamma\left(\frac{1}{5}\right)\Gamma\left(\frac{13}{10}\right)}\right)^{5/2}(\mu_R^{5/2}\pm\mu_L^{5/2}),\\
(J^b_{v/a})_z&=&-\frac{CB_x E_y}{a^6T^6L^6}
\left(\frac{3\sqrt{\pi}}{2\Gamma\left(\frac{1}{5}\right)\Gamma\left(\frac{13}{10}\right)}\right)^{5/2}(\mu_R^{5/2}\pm\mu_L^{5/2}),
\end{eqnarray}
where $a=4\pi/3$. 

On the contrary, in the limit of $\tilde{U}_{L/R}\rightarrow \infty$, which corresponds to low-density or high-temperature conditions, we find
\begin{eqnarray}
\frac{\mu_{L/R}}{U_T}\approx\frac{2}{3}\tilde{U}_{L/R}^{-5/2}-\frac{1}{13}\tilde{U}_{L/R}^{-15/2}
+\mathcal{O}(\tilde{U}_{L/R}^{-25/2}).
\end{eqnarray}
Up to the leading order in the expansion with respect to $\tilde{U}_{L/R}^{-1}$, we obtain
\begin{eqnarray}\label{Jtmurelation}
(J_{L/R})_t=\frac{3}{2}U_T^{3/2}\mu_{L/R}.
\end{eqnarray}
The boundary currents now read
\begin{eqnarray}\nonumber\label{JyJzhighT}
(J^b_{v/a})_y&=&\frac{CE_y}{2}\rho^2T^2L^{9/2}
\left(\left(1+\frac{9\mu_R^2}{8(a^2T^2L^3)^2}\right)\pm\left(1+\frac{9\mu_L^2}{8(a^2T^2L^3)^2}\right)\right)
\\
(J^b_{v/a})_z&=&-\frac{3CB_x E_y}{2a^3T^3L^{3/2}}(\mu_R\pm\mu_L).
\end{eqnarray}
One may further rewrite (\ref{JyJzhighT}) in terms of $\mu_V/\mu_A$,
\begin{eqnarray}\nonumber\label{Jbsmallmu}
(J^b_v)_y&=&CE_ya^2T^2L^{9/2}
\left(1+\frac{9}{8(a^2T^2L^3)^2}(\mu_V^2+\mu_A^2)\right),
\\\nonumber
(J^b_a)_y&=&\frac{9CE_y}{4a^2T^2L^{3/2}}\mu_V\mu_A,
\\
(J^b_{v/a})_z&=&-\frac{3CB_x E_y}{a^3T^3L^{3/2}}\mu_{V/A},
\end{eqnarray}
where $\mu_{V/A}=(\mu_R\pm\mu_L)/2$. The small-chemical-potential dependence here is consistent with that found in \cite{Huang:2013iia,Pu:2014cwa} and (\ref{paritysmallmu}).

In the presence of strong electromagnetic fields, we are unable to solve (\ref{solAt}) analytically with the strong-field approximation. Nevertheless, it is useful to further investigate the explicit dependence of the electromagnetic fields and charge densities for the boundary currents. When having large $E_y$ and finite $B_x$, we find $U_c^3\rightarrow L^3E_y^2$. By doing some algebra with (\ref{bdcurrents}), we obtain
\begin{eqnarray}\nonumber\label{JblargeE}
(J^b_v)_y&\approx& C L^{5/2}E_y^{5/2},
\\\nonumber
(J^b_a)_y&\approx& \frac{(J^b_v)_t(J^b_a)_t}{CL^{5/2}E_y^{8/3}},
\\
(J^b_{v/a})_z&\approx&-\frac{B_x(J^b_{v/a})_t}{E_y}.
\end{eqnarray}  
On the contrary, when having large $B_x$ and finite $E_y$, we find $U_c^3\rightarrow U_T^3$, which gives
\begin{eqnarray}\nonumber\label{JblargeB}
(J^b_v)_y&\approx& C \frac{U_T^{5/2}E_y}{B_x},
\\\nonumber
(J^b_a)_y&\approx& \frac{E_y(J^b_v)_t(J^b_a)_t}{CL^{3}B_x^{3}},
\\
(J^b_{v/a})_z&\approx&-\frac{E_y(J^b_{v/a})_t}{B_x}.
\end{eqnarray}  
\subsection{Numerical Results}
\begin{figure}[t]
\begin{minipage}{7cm}
\begin{center}
{\includegraphics[width=7.5cm,height=5cm,clip]{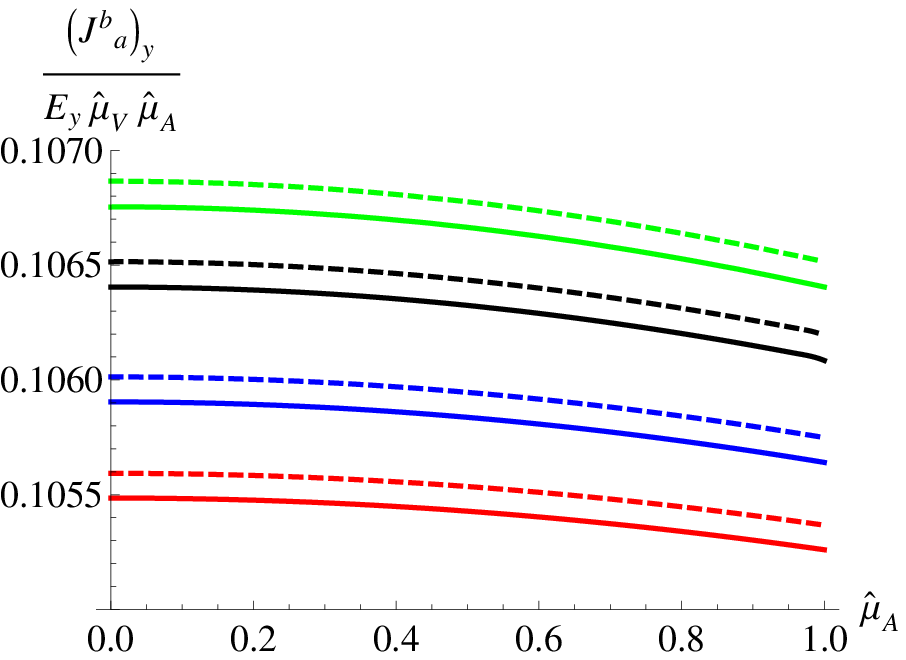}}
\caption{The green, black, blue, and red (from top to bottom) correspond to $\mu_V=0.002T$, $T$, $1.6T$, and $2T$, respectively. The solid and dashed curves correspond to $(B_x,E_y)=(m_{\pi}^2,m_{\pi}^2)=(0.135^2,0.135^2)$ GeV$^{2}$ and $(B_x,E_y)=(0.001^2,0.01^2)$ GeV$^{2}$, where $\hat{\mu}_{V/A}=\mu_{V/A}/T$.}\label{Jay_fixmuV}
\end{center}
\end{minipage}
\hspace {1cm}
\begin{minipage}{7cm}
\begin{center}
{\includegraphics[width=7.5cm,height=5cm,clip]{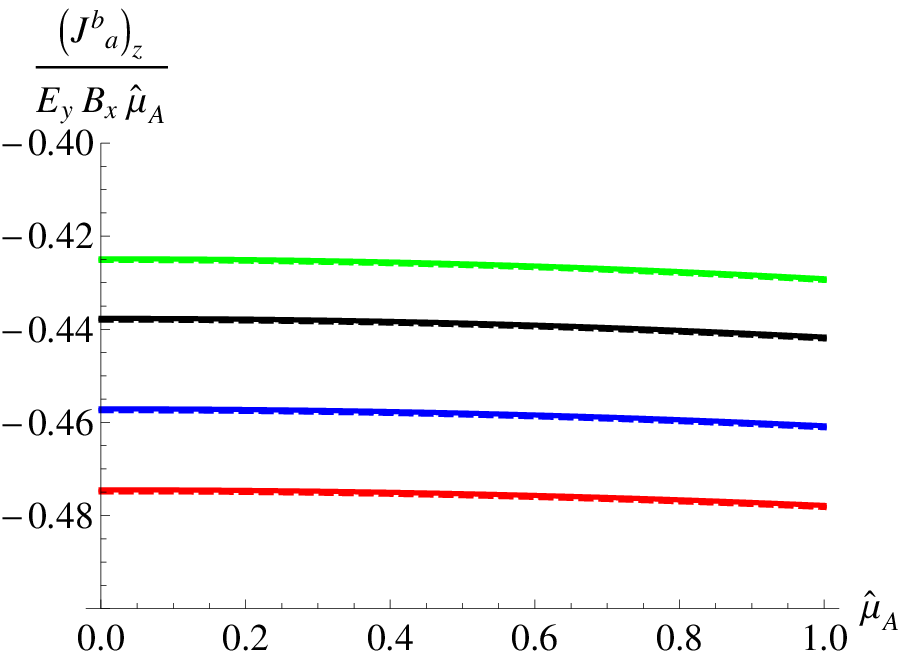}}
\caption{The color corresponds to the same cases in Fig.\ref{Jay_fixmuV}.}
\label{Jaz_fixmuV}
\end{center}
\end{minipage}
\end{figure}

\begin{figure}[t]
\begin{minipage}{7cm}
\begin{center}
{\includegraphics[width=7.5cm,height=5cm,clip]{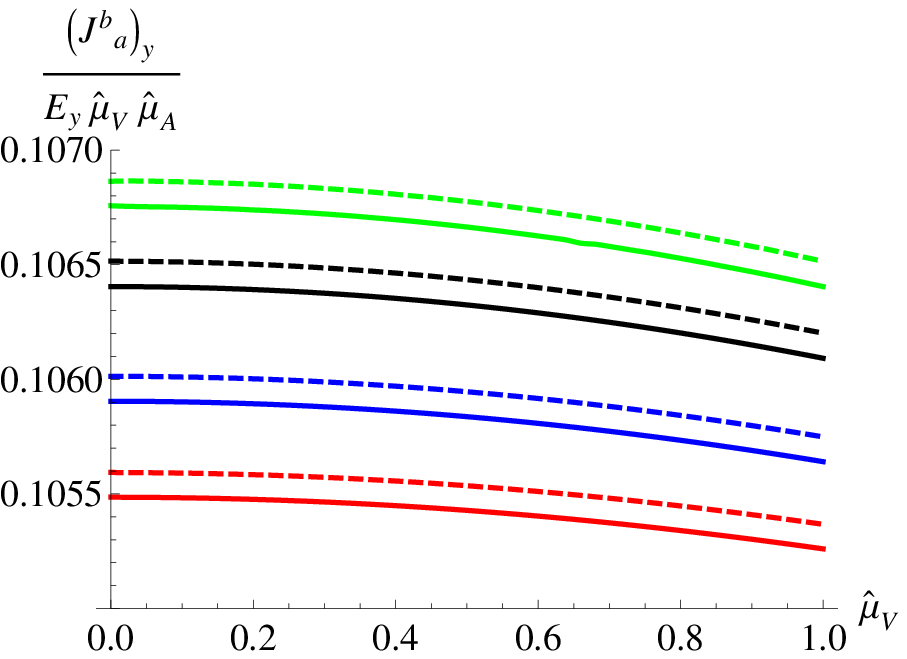}}
\caption{The green, black, blue, and red (from top to bottom) correspond to $\mu_A=0.002T$, $T$, $1.6T$, and $2T$, respectively. The solid and dashed curves correspond to $(B_x,E_y)=(m_{\pi}^2,m_{\pi}^2)=(0.135^2,0.135^2)$ GeV$^{2}$ and $(B_x,E_y)=(0.001^2,0.01^2)$ GeV$^{2}$.}\label{Jay_fixmuA}
\end{center}
\end{minipage}
\hspace {1cm}
\begin{minipage}{7cm}
\begin{center}
{\includegraphics[width=7.5cm,height=5cm,clip]{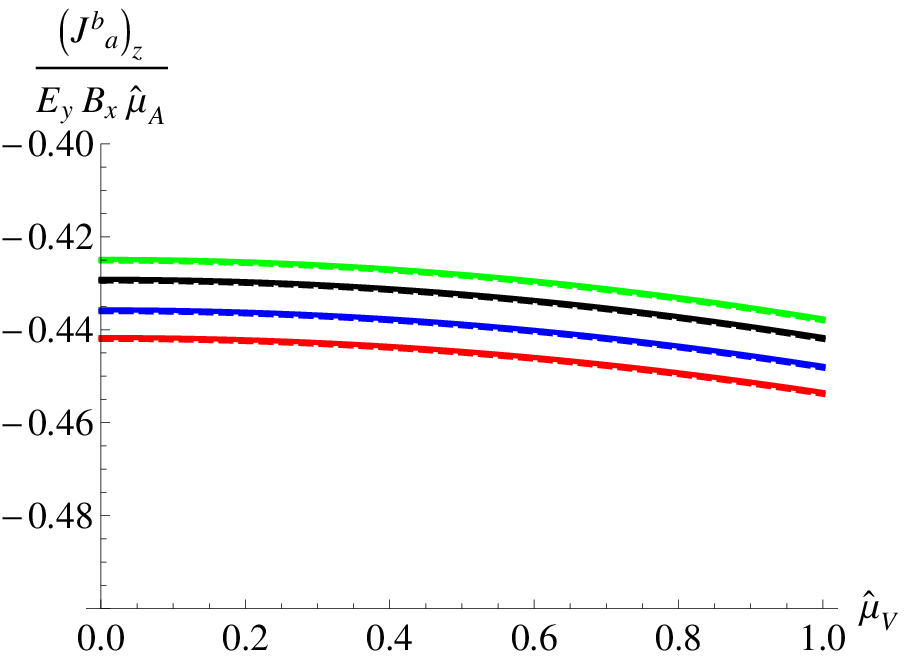}}
\caption{The color corresponds to the same cases in Fig.\ref{Jay_fixmuA}.}
\label{Jaz_fixmuA}
\end{center}
\end{minipage}
\end{figure}

\begin{figure}[t]
\begin{minipage}{7cm}
\begin{center}
{\includegraphics[width=7.5cm,height=5cm,clip]{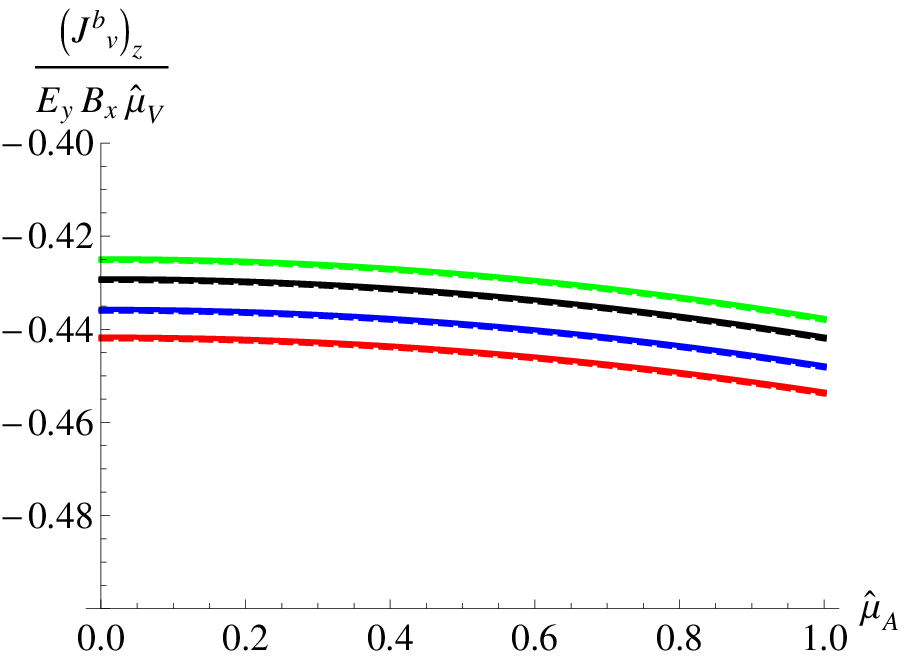}}
\caption{The color corresponds to the same cases in Fig.\ref{Jay_fixmuA}.}\label{Jvz_fixmuV}
\end{center}
\end{minipage}
\hspace {1cm}
\begin{minipage}{7cm}
\begin{center}
{\includegraphics[width=7.5cm,height=5cm,clip]{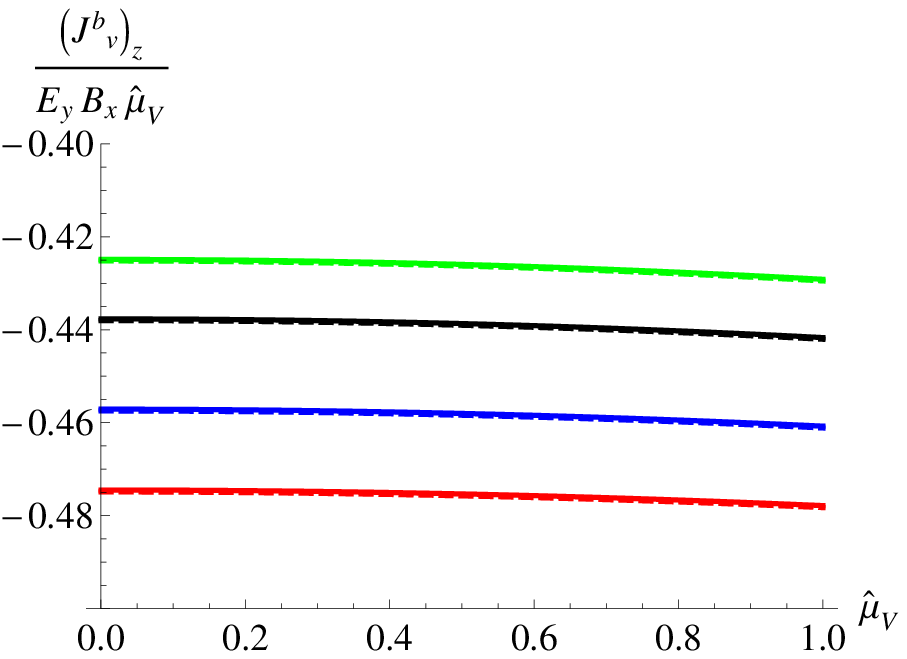}}
\caption{The color corresponds to the same cases in Fig.\ref{Jay_fixmuA}.}
\label{Jvz_fixmuA}
\end{center}
\end{minipage}
\end{figure}

We now numerically solve (\ref{solAt}) for the boundary currents.
The numerical values of the relevant coefficients are 
\begin{eqnarray}\label{parameter1}
2\pi l_s^2=1\text{GeV}^{-2},\quad \lambda_t=g_{YM}^2N_c=17,\quad M_{KK}=0.94\text{GeV}, 
\end{eqnarray}
which give
\begin{eqnarray}
L^3=(2M_{KK})^{-1}(g_{YM}^2N_cl_s^2)=1.44 \text{GeV}^{-3}.
\end{eqnarray}
We can further set $N_c=3$, which leads to $C=0.0211$ GeV$^{-15/2}$.
We then choose the temperature as the average temperature in RHIC,
\begin{eqnarray}
T=200\text{MeV}=0.2\text{GeV},
\end{eqnarray}
which yields
\begin{eqnarray}\label{parameter2}
U_T=1.02\text{GeV}^{-1}.
\end{eqnarray}
We firstly evaluate the axial currents generated by weak electromagnetic fields and by the average electromagnetic fields in RHIC\cite{Bzdak:2011yy,Hirono:2012rt} with different chemical potentials. In Fig.\ref{Jay_fixmuV} and Fig.\ref{Jaz_fixmuV}, we fix the vector chemical potentials and vary the axial chemical potentials by implementing the shooting method, where the currents are normalized by $C$. We find that the axial currents led by the CESE are approximately proportional to $\mu_V\mu_A$ even with finite chemical potentials. Our result is consistent with what have been found by using Kubo formula in \cite{Pu:2014cwa}. Moreover, the axial Hall currents are approximately linear to $\mu_A$, which match the approximation under weak electromagnetic fields and small chemical potentials. Analogously, the vector Hall currents are also approximately linear to $\mu_V$ as shown in Fig.\ref{Jvz_fixmuV}. 
It turns out that the small-chemical-potential approximation could be applied to the conditions when the chemical potentials are around the magnitude of the temperature. Also, the average electromagnetic fields in RHIC only result in minor corrections. The similar behaviors of the axial and vector currents can be found in Fig.\ref{Jay_fixmuA}, Fig.\ref{Jaz_fixmuA}, and Fig.\ref{Jvz_fixmuA} when we fix the axial chemical potentials and vary the vector ones.     

\begin{figure}[t]
     \begin{center}
        \subfigure[vector charge density]{%
            \label{fig:first}
            \includegraphics[height=4cm, width=0.4\textwidth]{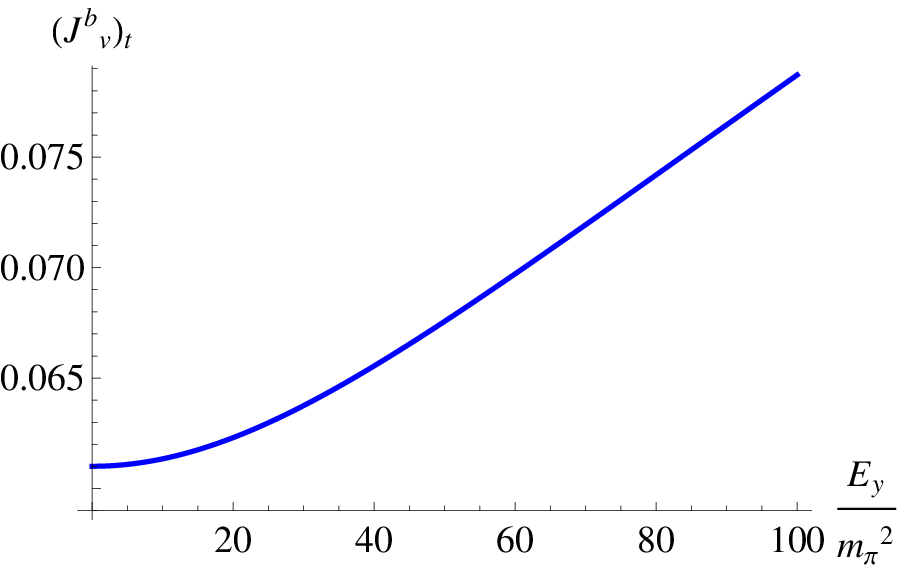}
        }%
        \subfigure[axial charge density]{%
           \label{fig:second}
           \includegraphics[height=4cm,width=0.4\textwidth]{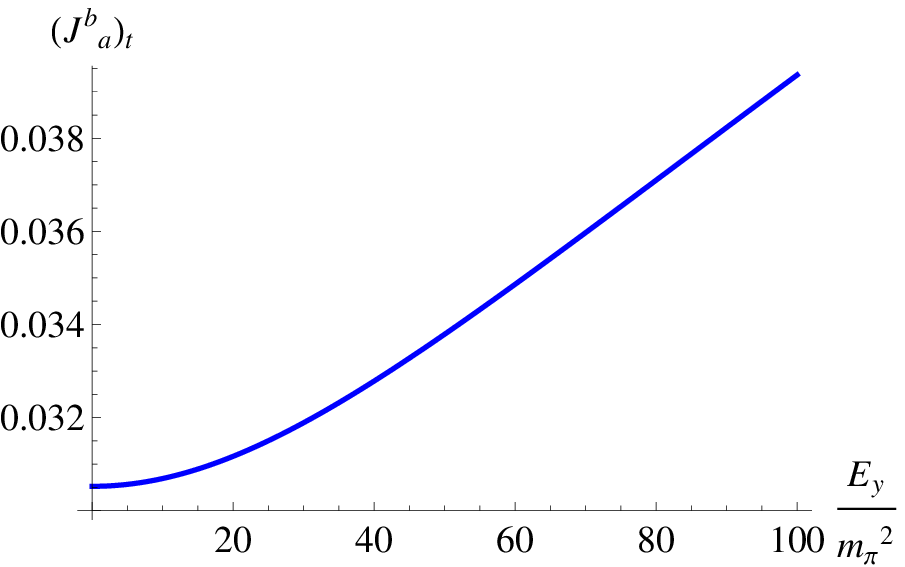}
        }\\ 
        \subfigure[vector current]{%
            \label{fig:third}
            \includegraphics[height=4cm,width=0.4\textwidth]{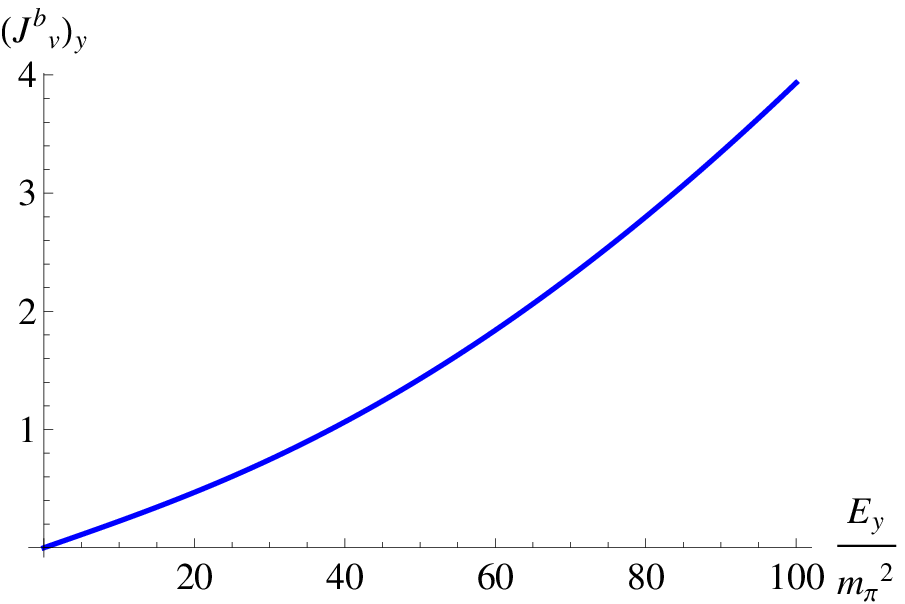}
        }%
        \subfigure[axial current]{%
            \label{fig:fourth}
            \includegraphics[height=4cm,width=0.4\textwidth]{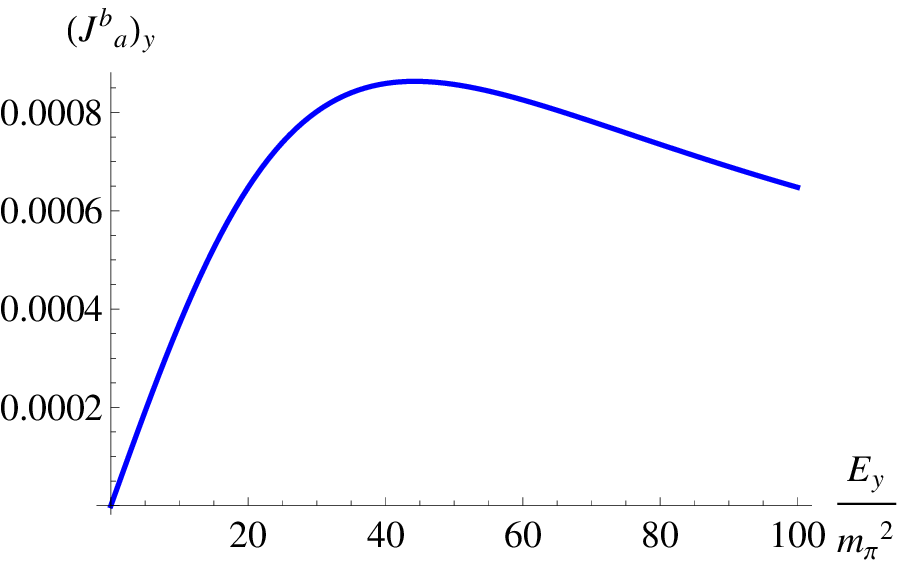}
        }\\
        \subfigure[vector Hall current]{%
            \label{fig:third}
            \includegraphics[height=4cm,width=0.4\textwidth]{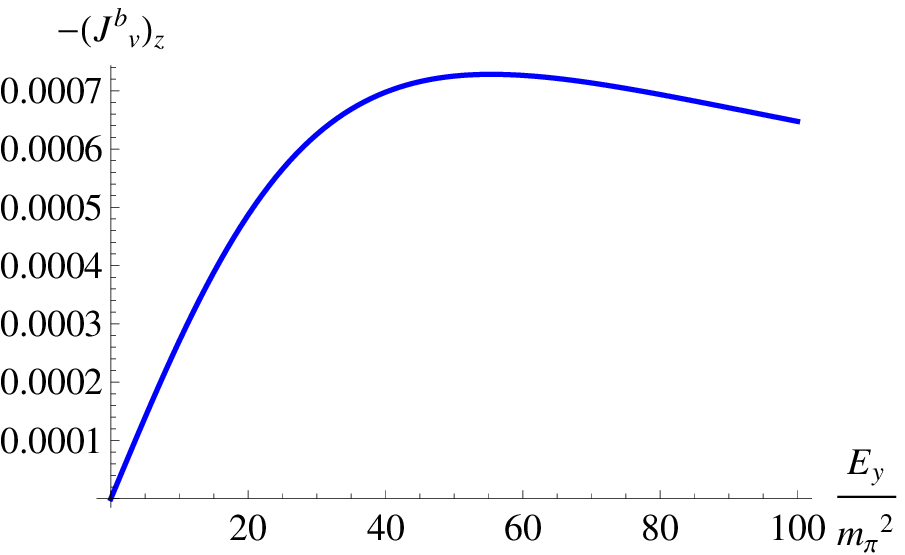}
        }%
        \subfigure[axial Hall current]{%
            \label{fig:fourth}
            \includegraphics[height=4cm,width=0.4\textwidth]{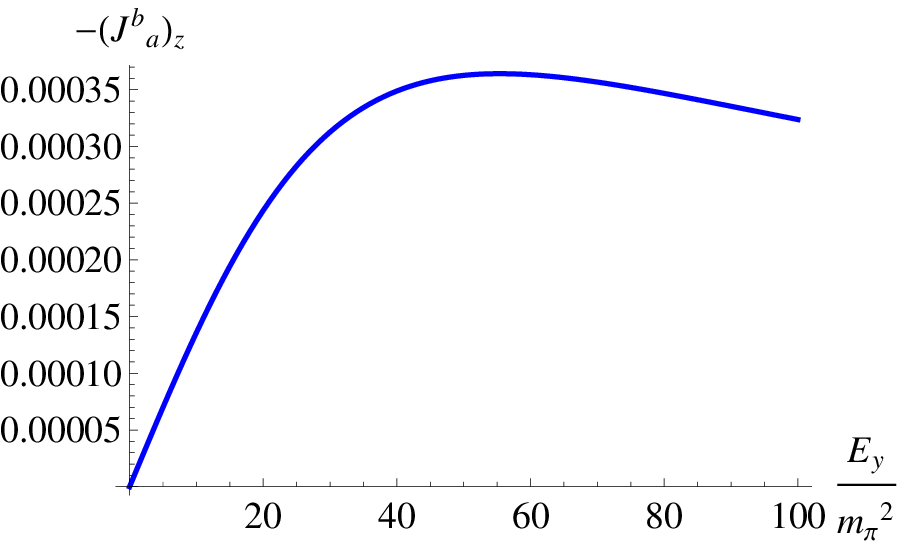}
        }
    \end{center}
    \caption{%
       Boundary currents normalized by $C$ with $B_x=m_{\pi}^2$, $\mu_V=0.2T$, and $\mu_A=0.1T$.
     }%
   \label{currentsfixB}
\end{figure}

\begin{figure}[t]
     \begin{center}
        \subfigure[vector charge density]{%
            \label{fig:first}
            \includegraphics[height=4cm,width=0.4\textwidth]{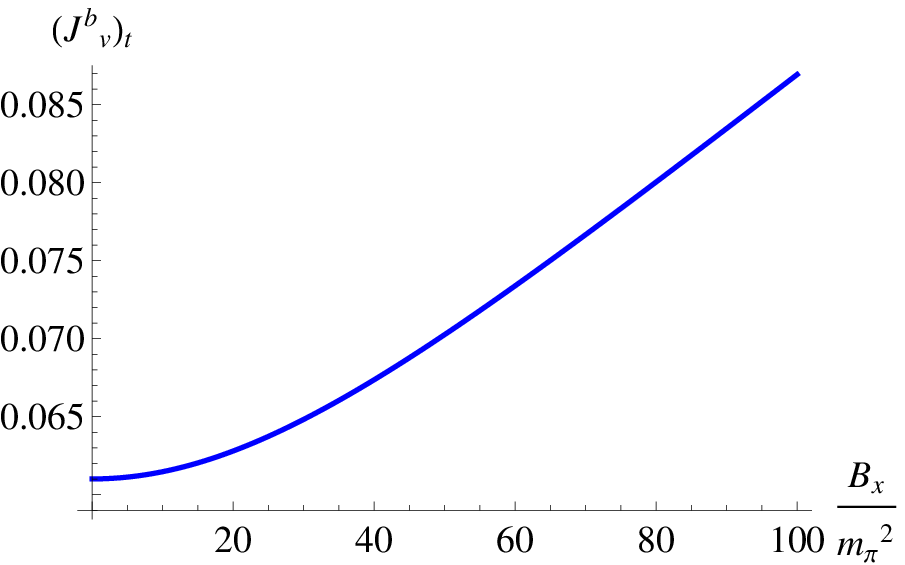}
        }%
        \subfigure[axial charge density]{%
           \label{fig:second}
           \includegraphics[height=4cm,width=0.4\textwidth]{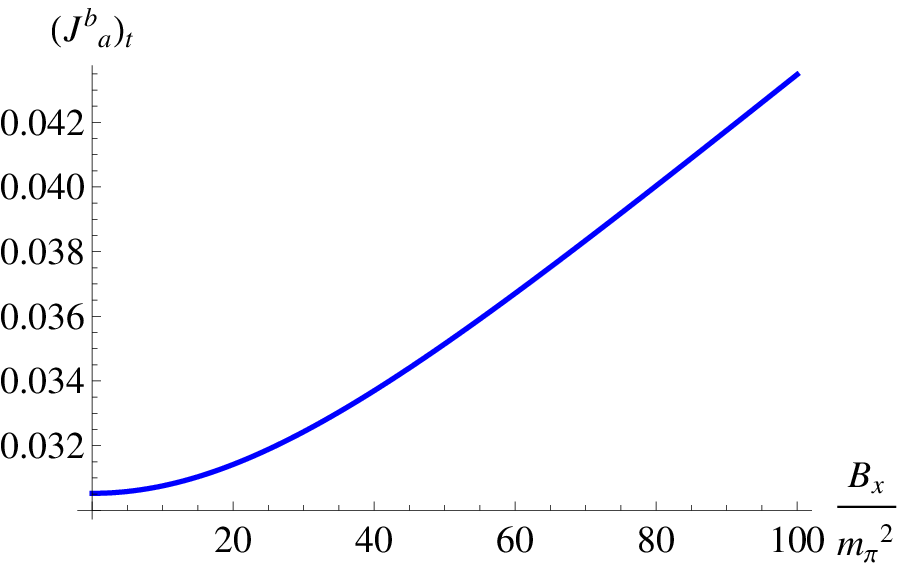}
        }\\ 
        \subfigure[vector current]{%
            \label{fig:third}
            \includegraphics[height=4cm,width=0.4\textwidth]{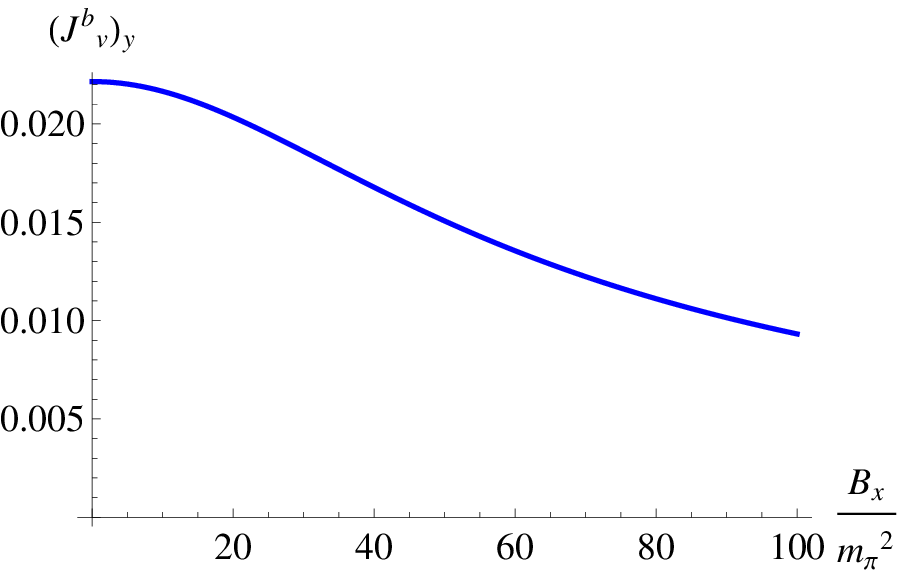}
        }%
        \subfigure[axial current]{%
            \label{fig:fourth}
            \includegraphics[height=4cm,width=0.4\textwidth]{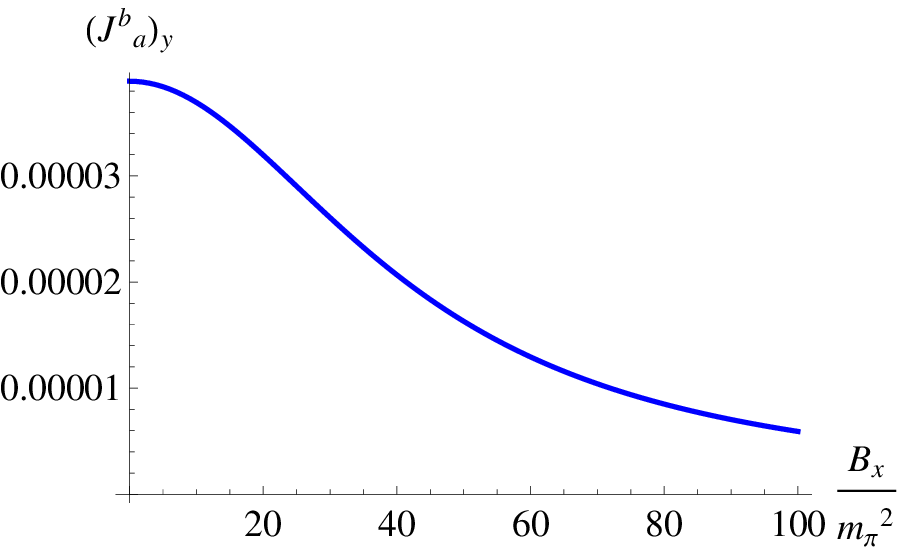}
        }\\
        \subfigure[vector Hall current]{%
            \label{fig:third}
            \includegraphics[height=4cm,width=0.4\textwidth]{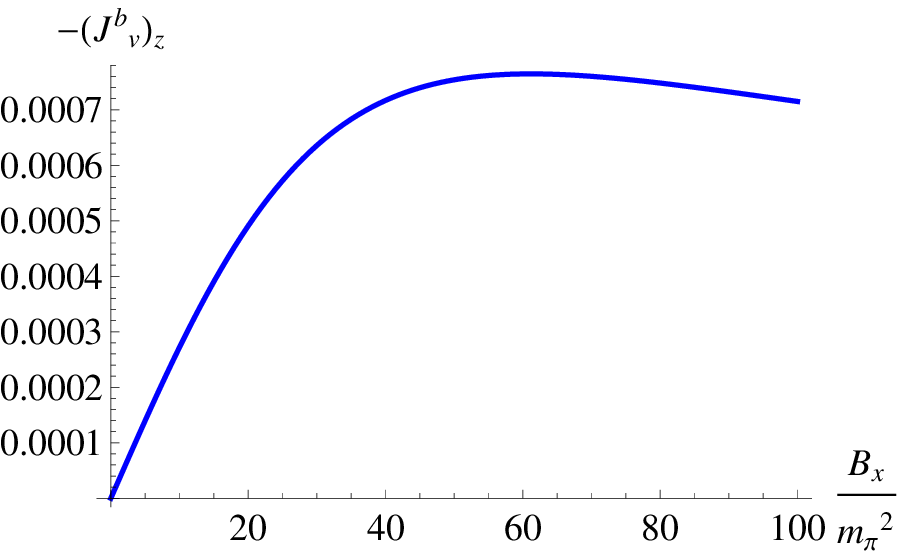}
        }%
        \subfigure[axial Hall current]{%
            \label{fig:fourth}
            \includegraphics[height=4cm,width=0.4\textwidth]{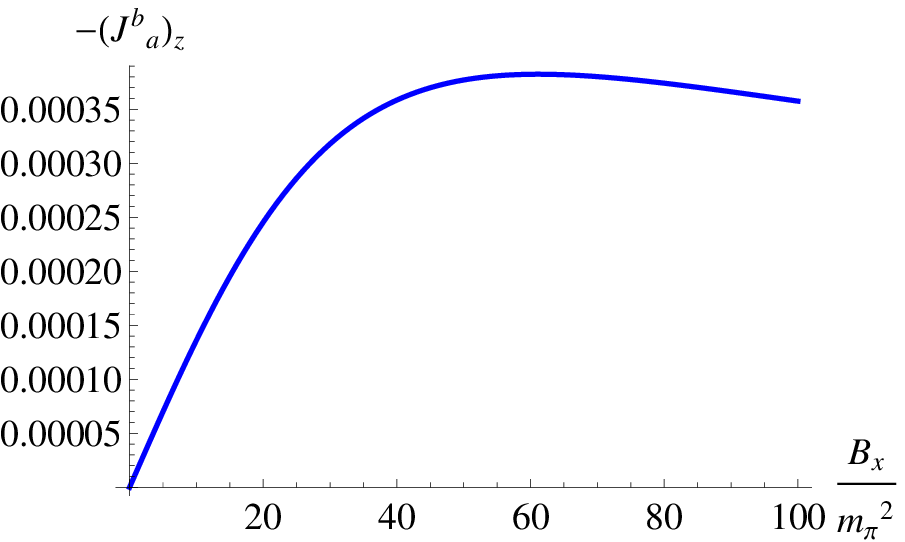}
        }
    \end{center}
    \caption{%
       Boundary currents normalized by $C$ with $E_y=m_{\pi}^2$, $\mu_V=0.2T$, and $\mu_A=0.1T$.
     }%
   \label{currentsfixE}
\end{figure}

Next, we may study the electric and Hall currents varied by electromagnetic fields. The numerical results are shown in Fig.\ref{currentsfixB}-\ref{currentsfixE}, where we fix both the vector and axial chemical potentials to be small compared with the temperature. In Fig.\ref{currentsfixB}, we fix $B_x$ to the average value in RHIC and vary $E_y$. In the regions of small electric field for $E_y<20m_{\pi}^2$, the increase of the charge densities led by $E_y$ is mild, while the currents $(J^b_{v/a})_{y/z}$ are linear to the electric field as expected from (\ref{Jbsmallmu}). In the region with a large $E_y$, the charge densities are increased by the electric field when fixing the chemical potentials, while the currents start to decrease except for $(J^b_v)_y$. The result could be qualitatively consistent with the strong-field approximation in (\ref{JblargeE}). However, the increase of $(J^b_{v/a})_t$ mitigates the decrease of $(J^b_a)_y$ and $(J^b_{v/a})_z$. In Fig.\ref{currentsfixE}, we then fix $E_y$ and vary $B_x$. We observe the linear increase of $(J^b_{v/a})_z$ as expected from (\ref{Jbsmallmu}). Also, the decrease of $(J^b_v)_y$ is mild with small $B_x$, but the nonlinear effect quickly takes over for $(J^b_a)_y$. In the region with large $B_x$, all currents decrease as anticipated from (\ref{JblargeB}).      

\section{CEW in holography}\label{CEW_holography}
\subsection{CEW in the SS Model}
In this section, we will investigate the transport coefficients of CEW in the frame work of the SS model. We may focus on the cases with weak electric fields such that the boundary currents are linear to the electric fields, while we may preserve the nonlinear effect from the magnetic fields encoded in the conductivities. Also, we will neglect the contributions from the CS terms.           
From (\ref{bcurrentRL}), we find
\begin{eqnarray}\nonumber\label{betayyzy}
(\beta_{L/R})_{yy}&=&\frac{L^{3/2}U_T^{3/2}}{B_x^2L^3+U_T^3}
\frac{(J_{L/R})_t}{\sqrt{(J_{L/R})_t^2+B_x^2L^3U_T^3+U_T^5}},\\
(\beta_{L/R})_{zy}&=&\frac{-B_xL^3}{B_x^2L^3+U_T^3},
\end{eqnarray}
where we take $U_c\approx U_T$ for small $E_y$. Since $(\beta_{L/R})_{zy}$ are independent of $(J_{L/R})_t$, we directly obtain $(\beta_-)_{zy}=0$ for arbitrary chemical potentials. We thus obtain
\begin{eqnarray}\nonumber
(\delta j_{v})_y&=&\left((\beta_+)_{yy}\delta j^0_{v}+(\beta_-)_{yy}\delta j^0_{a}\right)E_y,\\\nonumber
(\delta j_{a})_y&=&\left((\beta_-)_{yy}\delta j^0_{v}+(\beta_+)_{yy}\delta j^0_{a}\right)E_y,\\
(\delta j_{v/a})_z&=&(\beta_+)_{zy}\delta j^0_{v/a}E_y.
\end{eqnarray}
The transport coefficients in the dispersion relation read
\begin{eqnarray}\nonumber\label{transptcoeff}
\tau_{\pm}^{-1}&=&\left(n_v(\beta_+)_{yy}+\frac{(\sigma_v)_{yy}}{2}
\pm\sqrt{n_v^2(\beta_-)_{yy}^2+n_v(\beta_-)_{yy}(\sigma_a)_{yy}+\frac{(\sigma_v)_{yy}^2}{4}}\right),
\\\nonumber
(v_{\pm})_y&=&\left((\beta_{+})_{yy}\pm\frac{ (\beta_-)_{yy} (2 n_v (\beta_{-})_{yy}+(\sigma_a)_{yy})}{\sqrt{4n_v^2 (\beta_-)_{yy}^2+4 n_v(\beta_{-})_{yy}(\sigma_a)_{yy}+(\sigma_v)_{yy}^2}}\right)E_y,
\\\nonumber
(v_{\pm})_z&=&(\beta_{+})_{zy}E_y,
\\\nonumber
(D_{\pm})_{yy}&=&\mp\frac{(\beta_-)^2_{yy} \left((\sigma_v)_{yy}^2-(\sigma_a)_{yy}^2\right) (E_y^2)}{\left(4n_v^2 (\beta_-)_{yy}^2+4n_v(\beta_-)_{yy}(\sigma_a)_{yy}+(\sigma_v)_{yy}^2\right)^{3/2}},
\\
(D_{\pm})_{zz}&=&(D_{\pm})_{zy}=(D_{\pm})_{yz}=0.
\end{eqnarray}
Recall that $(\sigma_v)_{yy}>(\sigma_a)_{yy}$ in the limit of small chemical potentials. By further turning off $n_v$, we find that only the $\tau_-^{-1}$ vanishes. Therefore, when $n_v=0$, the dissipation of the "$-$" mode of CEW only comes from the diffusion. Although the diffusion constant for the "$+$" mode here is negative, the finite damping time should dominate the dissipation. The same argument can be applied to CMW showed in (\ref{CMWdispersion}) as well. Moreover, the "$-$" mode of the Hall CEW becomes non-dissipative when $n_v=0$ and $k_y=0$. This may be somewhat anticipated since the Hall currents are not influenced by the collisional effect in the "stationary state" in the absence of the currents along the electric field, which is equivalent to the condition with zero drag force or infinite relaxation time as discussed in Sec.\ref{sub:BE}.

\begin{figure}[t]
     \begin{center}
        \subfigure[]{%
            \label{rtimemfixEB}
            \includegraphics[height=4cm,width=0.4\textwidth]{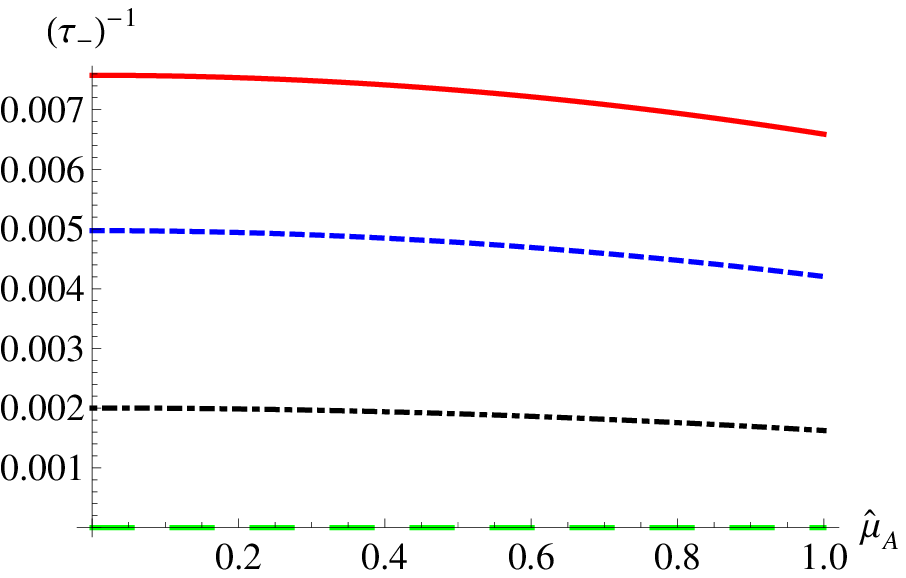}
        }%
        \subfigure[]{%
           \label{rtimepfixEB}
           \includegraphics[height=4cm,width=0.4\textwidth]{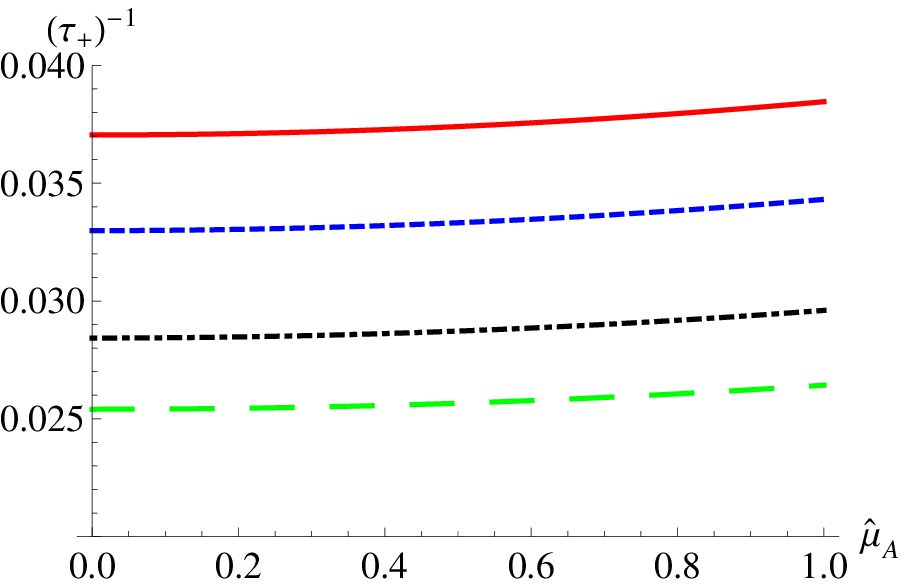}
        }\\ 
        \subfigure[]{%
            \label{vyfixEB}
            \includegraphics[height=4cm,width=0.4\textwidth]{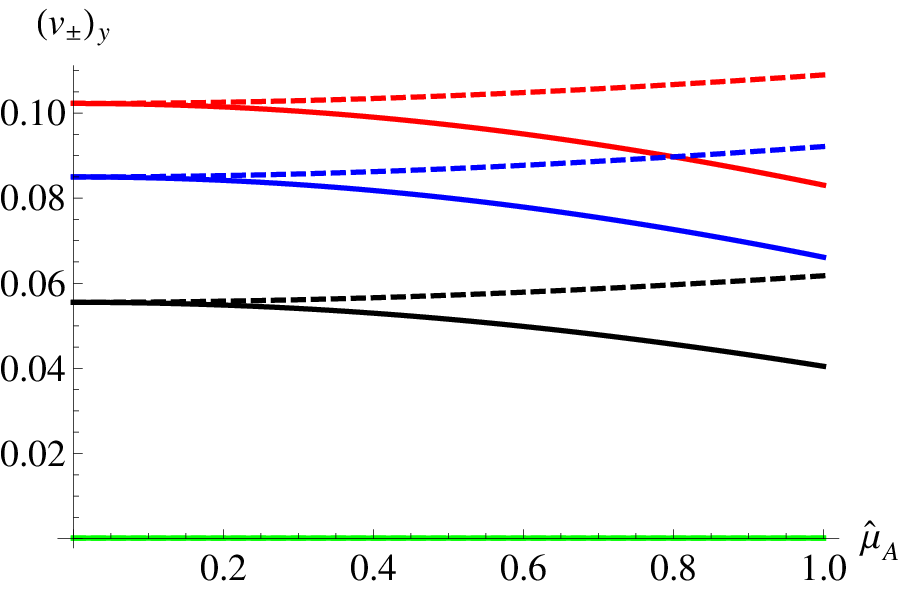}
        }%
        \subfigure[]{%
            \label{vyfixEB0mu}
            \includegraphics[height=4cm,width=0.4\textwidth]{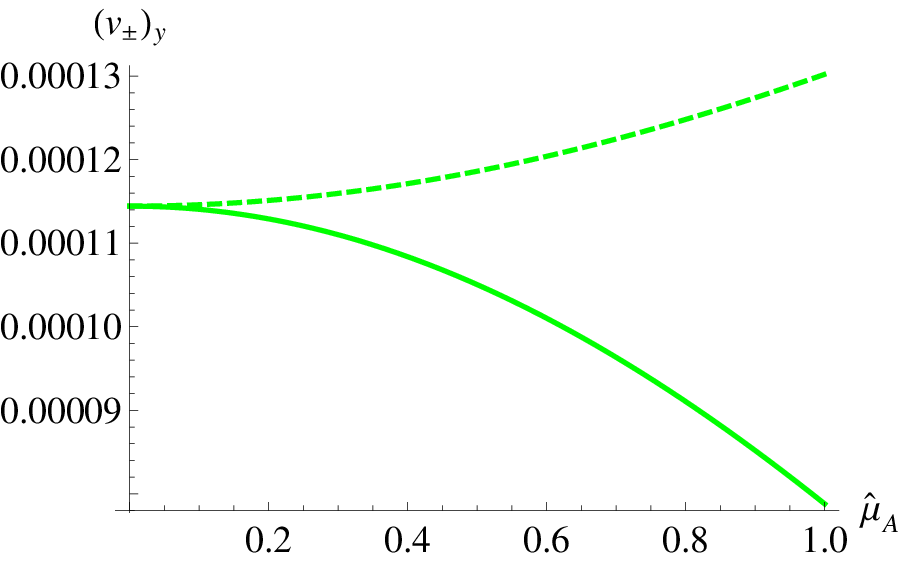}
        }\\
        \subfigure[]{%
            \label{vzfixEB}
            \includegraphics[height=4cm,width=0.4\textwidth]{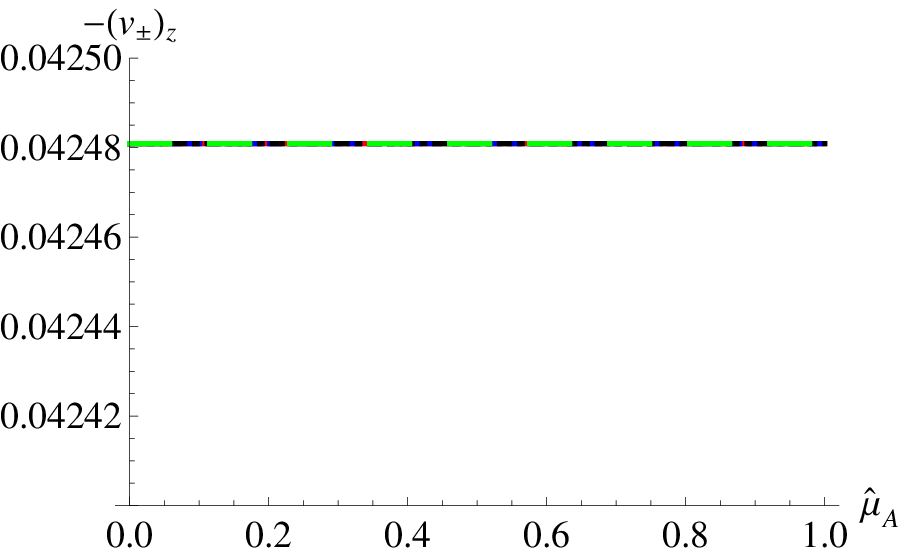}
        }%
        \subfigure[]{%
            \label{DyyfixEB}
            \includegraphics[height=4cm,width=0.4\textwidth]{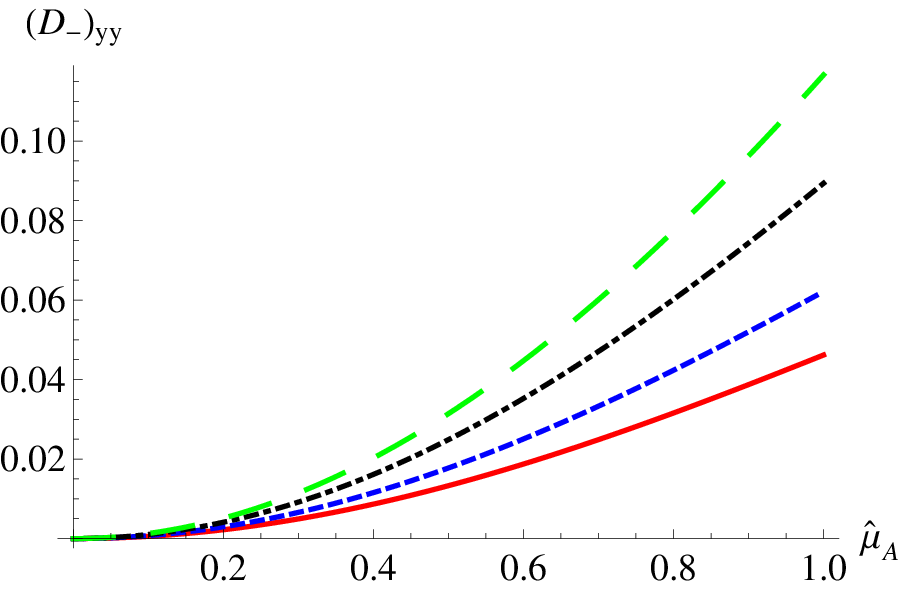}
        }
    \end{center}
    \caption{%
       The green, black, blue, and red (from bottom to top in (a)-(e) and from top to bottom in (f)) correspond to $\mu_V=0.002T$, $T$, $1.6T$, and $2T$, respectively. Here we take $E_y=B_x=10m_{\pi}^2$. The unit of $\tau_-$ is in GeV$^{-1}$. In (c), the solid and dashed curves represent $(v_-)_y$ and $(v_+)_y$.
     }%
\end{figure}

We now evaluate the transport coefficients in (\ref{transptcoeff}) numerically. We first consider the cases with fixed electromagnetic fields and different magnitudes of the chemical potentials. The results are shown in Fig.\ref{rtimemfixEB}-\ref{DyyfixEB}. As illustrated in Fig.\ref{rtimemfixEB} and Fig.\ref{rtimepfixEB}, the damping is more prominent for the "$+$" mode which mainly stems from the nonzero normal conductivity. For both two modes, the damping is increased by the vector chemical potential, while it is less affected by the axial chemical potential. Similarly, the wave velocities along the electric field of two modes are enhanced by the vector chemical potential and degenerate in the presence of an axial chemical potential as shown in Fig.\ref{vyfixEB} .
On the contrary, as expected from (\ref{betayyzy}) and (\ref{transptcoeff}), the Hall velocities of two modes as illustrated in Fig.\ref{vzfixEB} are degenerate and independent of the chemical potentials. As shown in Fig.\ref{DyyfixEB}, the diffusion constant vanishes at zero axial chemical potentials and increase when the axial chemical potential is increased. However, the diffusion constant is reduced by the vector chemical potential due to the presence of $n_v$ in the denominator as shown in (\ref{transptcoeff}).      

\begin{figure}[t]
     \begin{center}
        \subfigure[]{%
            \label{rtimemfixmuVT}
            \includegraphics[height=4cm,width=0.4\textwidth]{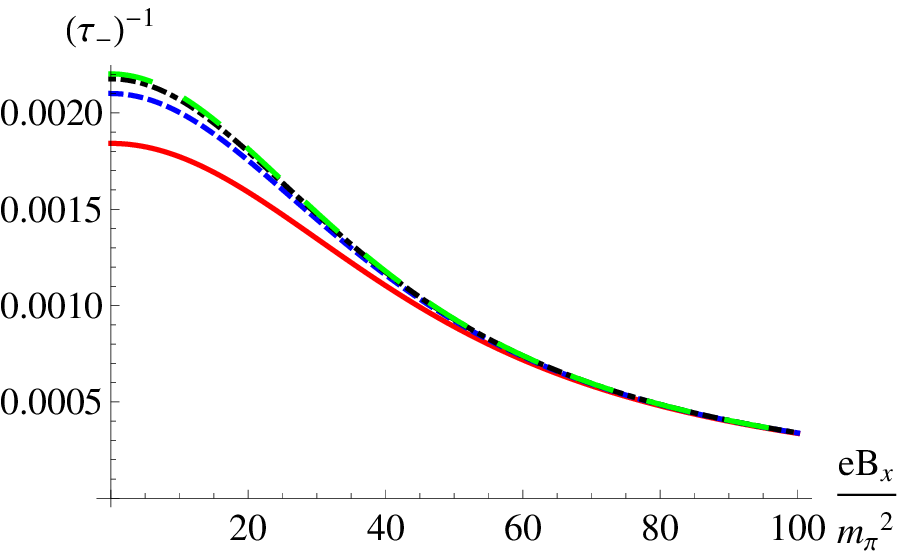}
        }%
        \subfigure[]{%
           \label{rtimepfixmuVT}
           \includegraphics[height=4cm,width=0.4\textwidth]{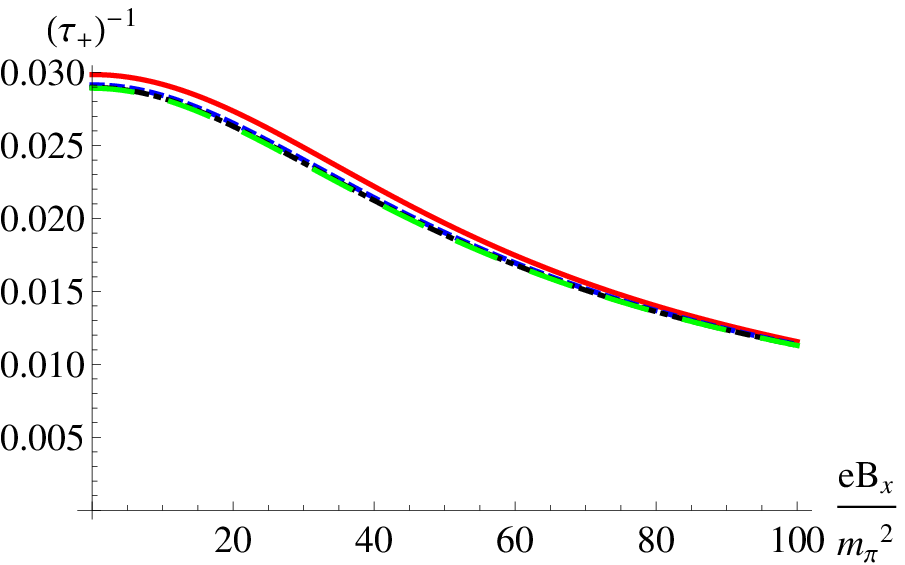}
        }\\ 
        \subfigure[]{%
            \label{vyfixmuVT}
            \includegraphics[height=4cm,width=0.4\textwidth]{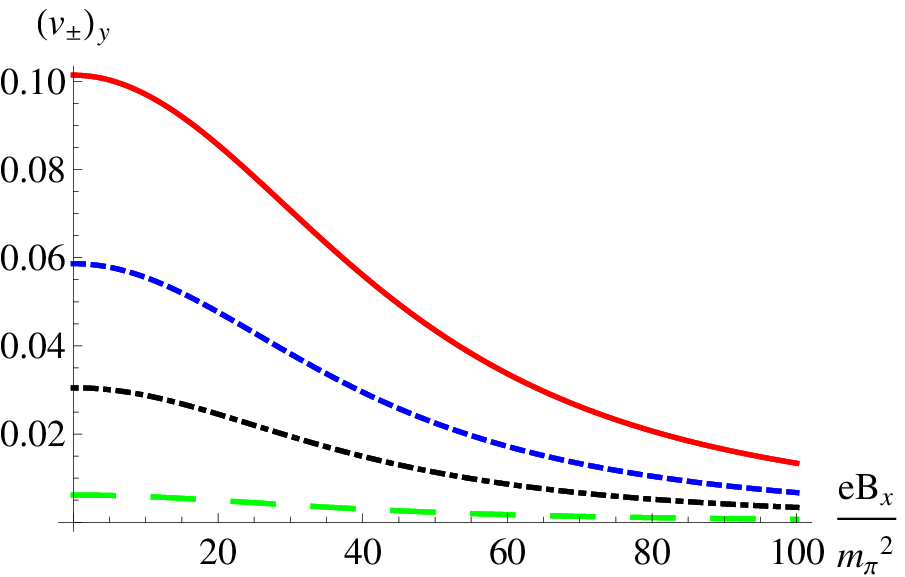}
        }%
        \subfigure[]{%
            \label{vzfixmuVT}
            \includegraphics[height=4cm,width=0.4\textwidth]{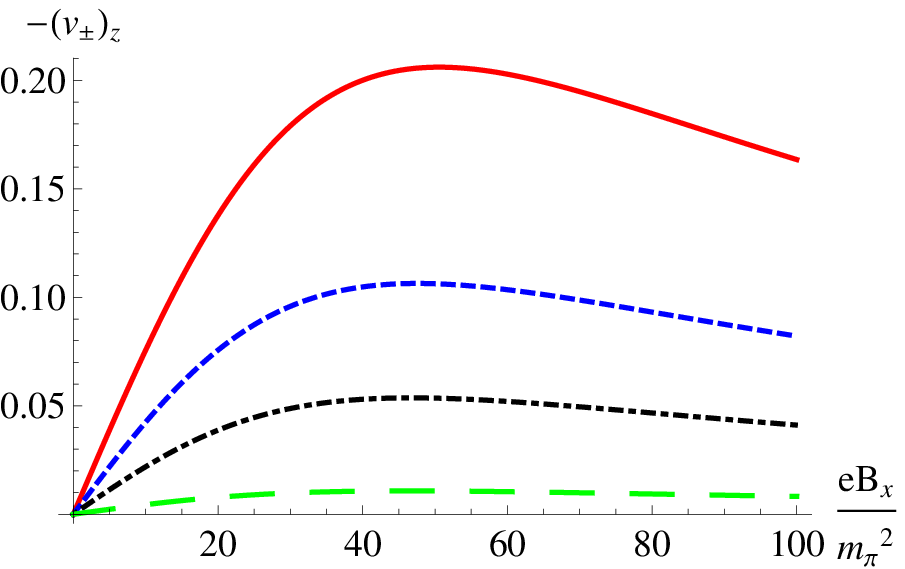}
        }
    \end{center}
    \caption{%
       The green(long-dashed), black(dot-dashed), blue(dashed), and red(solid) correspond to $eE_y=m_{\pi}^2$, $5m_{\pi}^2$, $10m_{\pi}^2$, and $20m_{\pi}^2$, respectively. Here we take $\mu_V=T$ and $\mu_A=0$. The unit of $\tau_-$ is in GeV$^{-1}$.
     }%
\end{figure}

\begin{figure}[t]
     \begin{center}
        \subfigure[]{%
            \label{rtimemfixmuV2T}
            \includegraphics[height=4cm,width=0.4\textwidth]{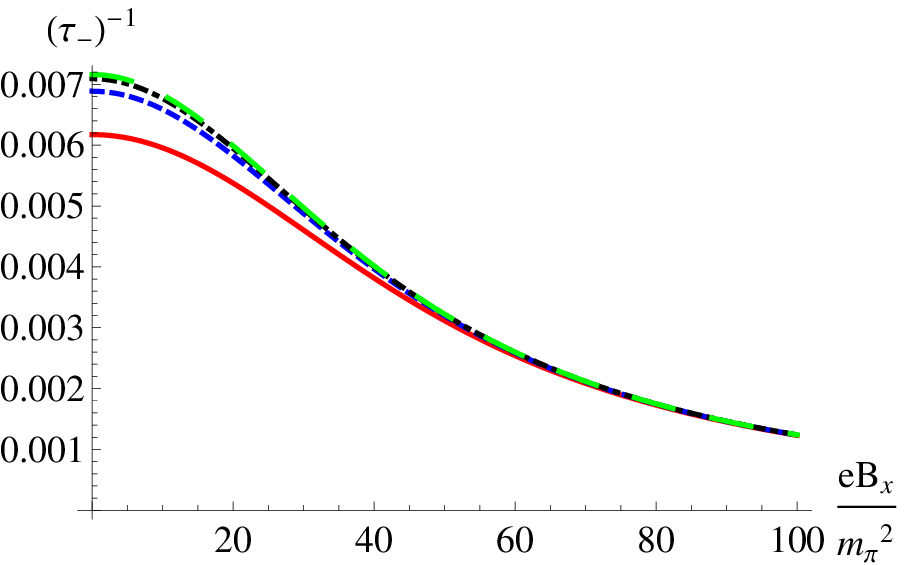}
        }%
        \subfigure[]{%
           \label{rtimepfixmuV2T}
           \includegraphics[height=4cm,width=0.4\textwidth]{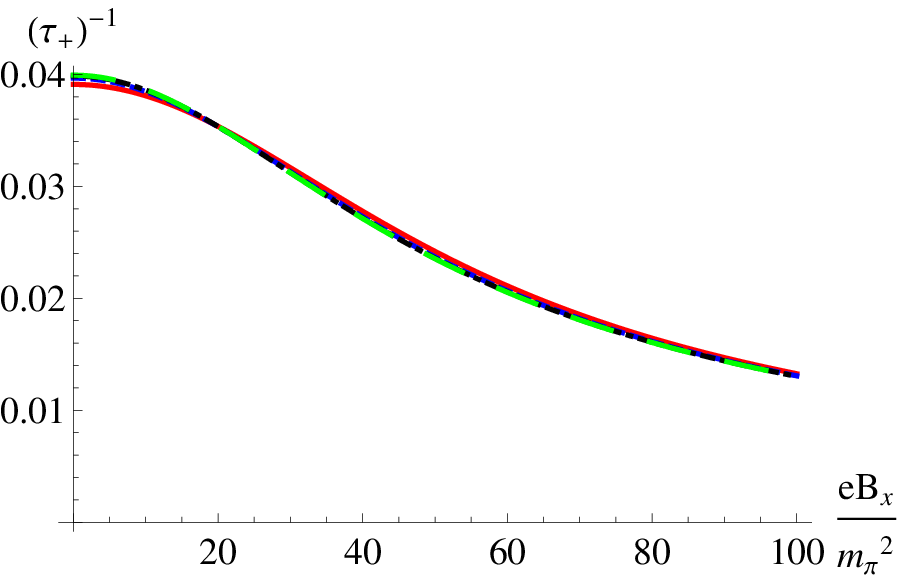}
        }\\ 
        \subfigure[]{%
            \label{vyfixmuV2T}
            \includegraphics[height=4cm,width=0.4\textwidth]{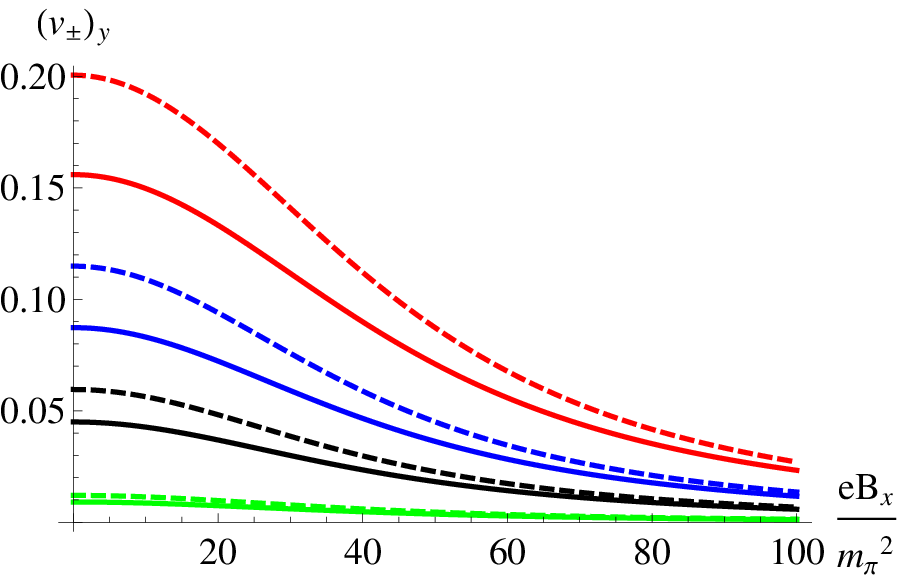}
        }%
        \subfigure[]{%
            \label{vzfixmuV2T}
            \includegraphics[height=4cm,width=0.4\textwidth]{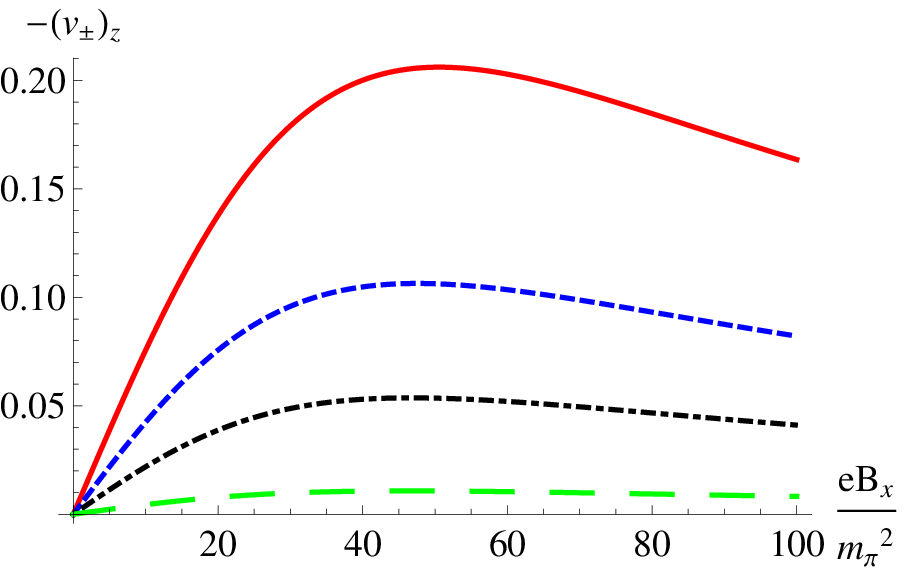}
        }\\
        \subfigure[]{%
            \label{DyyfixmuV2T}
            \includegraphics[height=4cm,width=0.4\textwidth]{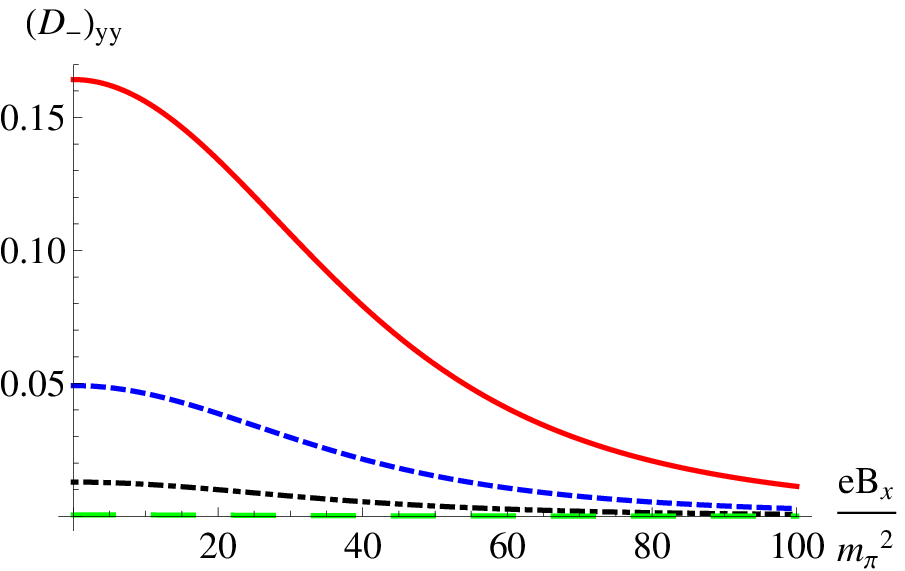}}
    \end{center}
    \caption{%
       The green(long-dashed), black(dot-dashed), blue(dashed), and red(solid) correspond to $eE_y=m_{\pi}^2$, $5m_{\pi}^2$, $10m_{\pi}^2$, and $20m_{\pi}^2$, respectively. Here we take $\mu_V=2T$ and $\mu_A=T$. The unit of $\tau_-$ is in GeV$^{-1}$. In \ref{vyfixEB}, the solid and dashed curves represent $(v_-)_y$ and $(v_+)_y$.
     }%
\end{figure}

Next, we may fix the chemical and vary the magnitudes of the constant electromagnetic fields. As shown in Fig.\ref{rtimemfixmuVT}-\ref{vzfixmuVT},
we plot the coefficients with $\mu_V=T$ and $\mu_A=0$. Since $(\beta_-)_{yy}=0$ when $\mu_A=0$, $(v_+)_y$ and $(v_-)_y$ are degenerate as illustrated in Fig.\ref{vyfixmuVT}. Also, $(D_\pm)_{yy}$ vanish under this condition. In Fig.\ref{rtimemfixmuV2T}-\ref{DyyfixmuV2T}, we take $\mu_V=2T$ and $\mu_A=T$, where the degeneracy of $(v_+)_y$ and $(v_-)_y$ is broken and $(D_{\pm})_{yy}$ are nonzero. Recall that $(D_+)_{yy}=-(D_-)_{yy}$. In addition, the magnitudes of $(D_{\pm})_{yy}$ will saturate to zero at large $B_x$, which could be expected from (\ref{transptcoeff}) since $(\beta_-)_{yy}$ drop to zero at large $B_x$ according to (\ref{betayyzy}). In general, when we increase the chemical potentials, the wave velocities increase, while the damping and diffusion contributing to the dissipation of CEW are enhanced as well. Nonetheless, with zero chemical potentials, the CEW may only propagate perpendicular to the applied fields without dissipation. Although the damping effect is absent only for the "$-$" mode here in the SS model due to presence of nonzero conductivity for the system at zero chemical potentials, both "$\pm$" modes for the Hall CEW will be non-dissipative in the system with zero conductivity and zero chemical potentials.    

\subsection{CEW in the Weakly/Strongly Coupled Scenarios at Small Chemical Potentials}
In this subsection, we may focus on the CEW at small chemical potentials in the absence of a magnetic field, where the transport coefficients for CEW can be derived analytically in both the SS model and weakly coupled QED through the conductivities obtained from the hard-thermal-loop approximation in \cite{Huang:2013iia}. 
For the weakly coupled scenario, we may consider an ideal gas at finite temperature and chemical potentials. The bookkeeping result(also see, for example, the number density for massless particles quarks in QGP in \cite{Pu:2011vr}) shows that
\begin{eqnarray}
j^0_{R/L}=\frac{Q_f\mu_{R/L}}{6}\left(T^2+\frac{\mu_{R/L}^2}{\pi^2}\right),
\end{eqnarray}
which results in 
\begin{eqnarray}
\alpha_{R/L}=\frac{6}{Q_fT^2\left(1+\frac{3\mu_{R/L}^2}{\pi^2T^2}\right)}
\approx\frac{6}{Q_fT^2}\left(1-\frac{3\mu_{R/L}^2}{\pi^2T^2}\right)
\end{eqnarray}
for small chemical potentials, where $Q_f$ denotes the degrees of freedom of the chiral fermions.
By definition, we find
\begin{eqnarray}
\beta_{R/L}=2\rho\mu_{R/L}\alpha_{R/L}
\approx\frac{12\tilde{\rho}\mu_{R/L}}{Q_fT^3}\left(1-\frac{3\mu_{R/L}^2}{\pi^2T^2}\right),
\end{eqnarray}
where $\tilde{\rho}=\rho T$ is dimensionless. We thus have
\begin{eqnarray}
\beta_{+/-}=\frac{12\tilde{\rho}}{Q_fT^3}\mu_{V/A}+\mathcal{O}(\mu_{R/L}^3/T^3).
\end{eqnarray}
In the limit $\mu_R=-\mu_L=\mu_A$ and $\sigma_{v/a}=0$, from (\ref{dispCMW1}), the dispersion relation for CMW reads
\begin{eqnarray}
\omega_{\pm}=\pm\lambda({\bf B\cdot k})\alpha_{R/L}=\pm\frac{eN_c{\bf B\cdot k}}{2\pi^2T^2}\left(\frac{6}{Q_f}\right)\left(1-\frac{3\mu_A^2}{\pi^2T^2}\right).
\end{eqnarray}
Analogously, from (\ref{dispCEW1}), the dispersion relation for CEW is given by
\begin{eqnarray}
\omega_{\pm}=\pm\lambda({\bf E\cdot k})\beta_{R/L}=\pm e{\bf E\cdot k}\frac{2\tilde{\rho}\mu_A}{T^3}\left(\frac{6}{Q_f}\right)\left(1-\frac{3\mu_A^2}{\pi^2T^2}\right).
\end{eqnarray}

The numerical value of $\tilde{\rho}$ depends on the property of the medium. In the weakly coupled QED, one can read out $\sigma_0$ and $\tilde{\rho}$ defined in (\ref{JRLsmallmu}) from \cite{Huang:2013iia} by turning off the contributions from the interaction between the right-handed and left-handed sectors
\footnote{We simply drop the terms proportional to $\mu_R^2+\mu_L^2$ in $\sigma_{R/L}$ in \cite{Huang:2013iia}. One may choose an alternative way to truncate the interaction by dropping the term proportional to $\mu_R^2(\mu_L^2)$ in $\sigma_L(\sigma_R)$. In such a case, we have $\tilde{\rho}\approx 9.005/(e^4\ln (1/e))$.}
, where 
\begin{eqnarray}
\sigma_0=15.6952\frac{T}{e^4\ln(1/e)},\quad\tilde{\rho}=10.2495\frac{1}{e^4\ln(1/e)}.
\end{eqnarray} 
Here we may consider two particular cases for CEW. When $n_v=0(\beta_+=0)$, from (\ref{CEWomega}) and (\ref{betapm}),
we find
\begin{eqnarray}
\omega_{\pm}\approx\pm e\sqrt{\beta_-^2E_y^2k_y^2-\frac{\sigma_v^2}{4}}-\frac{ie\sigma_v}{2}
\approx\pm e\sqrt{\left(\frac{6\tilde{\rho}\mu_AE_y}{T^3}\right)^2k_y^2-\frac{\sigma_0^2}{4}}-\frac{ie\sigma_0}{2},
\end{eqnarray}
where the contribution from $\sigma_A$ is dropped since $\sigma_A\sim\mathcal{O}(n_{R/L}^2)$. Here we take $Q_f=2$ by summing over the spins of electrons in QED.
The small-momentum expansions of two modes up to the leading order of $k_y$ are
\begin{eqnarray}\nonumber\label{CEWdispnv0}
\omega_+&=&-ie\left(\frac{6\tilde{\rho}\mu_AE_y}{T^3}\right)^2\frac{k_y^2}{\sigma_0}
=-i\frac{240.957(eE_y)^2\mu_A^2}{e^5\ln(1/e)T^7}k_y^2,
\\
\omega_-&=&-ie\sigma_0+ie\left(\frac{6\tilde{\rho}\mu_AE_y}{T^3}\right)^2\frac{k_y^2}{\sigma_0}
=-\frac{iT}{e^3\ln(1/e)}\left(15.6952-\frac{240.957(eE_y)^2\mu_A^2}{e^2T^8}k_y^2\right),
\end{eqnarray}
where both modes do not propagate. 
On the other hand, when $n_a=0(\beta_-=0)$, we have  
\begin{eqnarray}\nonumber\label{CEWdispna0}
\omega_{+}&=&e\left(\frac{6\tilde{\rho}\mu_VE_y}{T^3}\right)k_y
=\frac{61.4969\mu_V(eE_y)}{e^4\ln(1/e)T^3}k_y
\\
\omega_{-}&=&e\left(\frac{6\tilde{\rho}\mu_VE_y}{T^3}\right)k_y
-i e\sigma_0
=\frac{61.4969\mu_V(eE_y)}{e^4\ln(1/e)T^3}k_y-i\frac{15.9652}{e^3\ln(1/e)},
\end{eqnarray}
where we drop $n_v\beta_+\sim\mathcal{O}(n_{R/L}^2)$. In \cite{Huang:2013iia}, the interaction between the right-handed and left handed fermions was included. When $n_V=0$, in our convention, the dispersion relation of the CEW reads
\begin{eqnarray}\label{Huangnv0}
\omega_{\pm}=\pm e\sqrt{(v_ek_y^2)-(\sigma_0/2)^2}-ie\sigma_0/2,\quad v_e=\alpha_An_a\sqrt{2\sigma_2\chi_e\alpha_V\alpha_A}E_y,
\end{eqnarray}
where
\begin{eqnarray}
\sigma_2=7.76052\frac{1}{Te^4\ln(1/e)},\quad\chi_e=20.499\frac{1}{Te^4\ln(1/e)},
\quad\alpha_{V/A}=\frac{\partial\mu_{V/A}}{\partial j^0_{V/A}}\approx\frac{3}{T^2}.
\end{eqnarray}
By making the small-momentum expansion, (\ref{Huangnv0}) becomes
\begin{eqnarray}\nonumber
\omega_+&=&-ie\frac{v_e^2k_y^2}{\sigma_0}=-i\frac{182.444(eE_y)^2\mu_A^2}{e^5\ln(1/e)T^7}k_y^2,
\\
\omega_-&=&-ie\sigma_0+ie\frac{v_e^2k_y^2}{\sigma_0}
=-\frac{iT}{e^3\ln(1/e)}\left(15.6952-\frac{182.444(eE_y)^2\mu_A^2}{e^2T^8}k_y^2\right),
\end{eqnarray}
where the diffusion is enhanced by the interaction between the R/L sectors.
When $n_a=0$, two modes read
\begin{eqnarray}\nonumber\label{Huangna0}
\omega_{+}&=&ev_ak_y
=\frac{61.4969\mu_V(eE_y)}{e^4\ln(1/e)T^3}k_y
\\
\omega_{-}&=&ev_vk_y
-i e\sigma_0
=\frac{46.5631\mu_V(eE_y)}{e^4\ln(1/e)T^3}k_y-i\frac{15.9652T}{e^3\ln(1/e)},
\end{eqnarray}
where
\begin{eqnarray}
v_a=\chi_e\alpha_V\alpha_An_vE_y,\quad v_v=2\sigma_2\alpha_V^2n_vE_y.
\end{eqnarray}
Similar to (\ref{CEWdispna0}), the $\omega_{-}$ mode will be damped out but the velocities of these two modes are different in (\ref{Huangna0}) due to the interactions between the R/L sectors. When turning off the interactions, two velocities become degenerate. In \cite{Huang:2013iia}, the $\omega_-$ and $\omega_+$ modes are called the "vector density wave" and the "axial density wave", respectively. Here we find that only the axial density wave is unaffected by the interaction.

We may compare the results obtained from weakly coupled QED with that found in strongly coupled QCD(SS model). From (\ref{bdcurrents}) and (\ref{Jtmurelation}), we find
\begin{eqnarray}
\beta_{+/-}=\frac{3\mu_{V/A}(eE_y)}{2a^5T^5L^6}(2\pi l_s^2)^2,
\quad\sigma_v=Ca^2T^2L^{9/2}(2\pi l_s^2)^2,
\end{eqnarray}
where we write out the dependence of $2\pi l_s^2$ explicitly for dimensional analysis. When $n_v=0$, we have 
\begin{eqnarray}\nonumber
\omega_+&=&-i\frac{9(2\pi l_s^2)^2(eE_y)^2\mu_A^2}{4Ca^{12}L^{33/2}T^{12}}k_y^2,
\\
\omega_-&=&-i(2\pi l_s^2)^2\left(Ca^2T^2L^{9/2}+\frac{9(eE_y)^2\mu_A^2}{4Ca^{12}L^{33/2}T^{12}}k_y^2\right).
\end{eqnarray}
When $n_a=0$, we have
\begin{eqnarray}\nonumber\label{omegana0}
\omega_+&=&\frac{3\mu_V(2\pi l_s^2)^2(eE_y)k_y}{2a^5T^5L^6},
\\
\omega_-&=&(2\pi l_s^2)^2\left(\frac{3\mu_V(eE_y)k_y}{2a^5T^5L^6}
-iCa^2T^2L^{9/2}\right).
\end{eqnarray}
It turns out that the CEW in weakly coupled and in strongly coupled systems have different temperature dependence. In the weakly coupled QED, the hard-thermal-loop approximation assume that the temperature dominates all other scales in the system. However, the SS model contains $M_{KK}$ corresponding to the mesonic scale, which should be also involved in CEW. We may now focus on the propagating waves for $n_v=0$. By using $L^3=(4\pi M_{KK})^{-1}\lambda_t$ and 
$C=(12\pi^2L^{3/2})^{-1}N_c$ from $2\pi l_s^2=1$ GeV$^{-2}$, (\ref{omegana0}) can be written as
\begin{eqnarray}\nonumber
\omega_+&=&\frac{729M_{KK}^2}{128\pi^2\lambda_t^2T^2}\frac{(eE_y)\mu_V}{T^3}k_y,
\\
\omega_-&=&\frac{729M_{KK}^2}{128\pi^2\lambda_t^2T^2}\frac{(eE_y)\mu_V}{T^3}k_y
-i\frac{2\lambda_tN_cT^2}{54\pi M_{KK}}.
\end{eqnarray}    
In comparison with (\ref{Huangna0}), the diffusion constants for $\omega_-$ in the weakly coupled and strongly coupled scenarios have distinct dependence of both the temperature and coupling constants. 

\section{Summary and Outlook}\label{sum_outlook}  
In this work, we have proposed the chiral Hall effect(CHE) generated by the applied electromagnetic fields and an axial chemical potential. In the presence of an electric field and a magnetic field perpendicular to each other, collective excitations of thermal plasmas with nonzero vector and axial chemical potentials will result in density waves as the chiral electric waves(CEW) propagating along the directions parallel to the electric field and perpendicular to both applied fields. Although the CEW induced by the CESE only exist with nonzero chemical potentials, the CEW led by the CHE should survive even at zero chemical potentials. Such Hall CEW become non-dissipative at zero conductivity. In phenomenology, we have argued that the CHE could lead to rapidity-dependent charge asymmetry in asymmetric heavy ion collisions. Combining with the CME and CESE, we may find different charge asymmetry of flow harmonics $v_n$ at distinct rapidity.

Nevertheless, we are unable to draw the conclusion upon the magnitudes of the charge asymmetry of $v_n$ since the axial chemical potential in the QGP is unknown. Moreover, to describe the practical condition in heavy ion collisions, numerical simulations based on the wave equations derived in our work with proper initial charge distributions and hydrodynamic evolution of the QGP are needed. On the other hand, the topological effect in the QGP could be pronounced, we thus have to couple CEW with CMW. Also, in our work, we only consider the density fluctuations and neglect the fluctuation of the induced electromagnetic fields.
It has been indicated in \cite{Akamatsu:2013pjd,Akamatsu:2014yza} that the induced electromagnetic fields could further cause chiral-plasma instabilities in the presence of an external magnetic field. Such instabilities will reduce the CME. Therefore, it is tentative to explore the existence of similar instabilities for CESE and CHE in the future.

In holography, a substantial problem occurs when we try to compute the all currents generated by CME, CESE, and CHE, where the currents are not gauge invariant when incorporating the contributions from the CS terms in the SS model. 
Moreover, there exists a persistent debate upon the presence of the CME in the SS model, where the CME current cannot be both conserved and gauge-invariant. On the other hand, in \cite{Hoyos:2011us}, the CME is reproduced in holography via a different definition of the axial chemical potential in the D3/D7 system, where the axial chemical potential comes from the rotating D7 branes instead of the temporal gauge fields in the gravity dual. It is thus intriguing to investigate the CESE and CHE along with the CME in the frame work of the D3/D7 system. 

Furthermore, the Hall and chiral Hall effects can still survive
in non-relativistic systems, e.g. Weyl semi-metal. Quite different
from the spin Hall effect in the Weyl semi-metal induced by axion fields
or Berry phase, the chiral Hall effect in our work is caused by interactions, which 
will play a role if there is an effective $\mu_{A}$. We will
leave the applications to condensed matter system in the future.     

\section{Acknowledgment}
The authors thank Jiunn-wei Chen and Xu-guang Huang for helpful
discussions and valuable comments from the referee of Physical Review D. This work was supported by the NSFC under grant No.
11205150. SP was supported in part by the NSC, NTU-CTS, and the
NTU-CASTS of R.O.C. SYW was supported
by the National Science Council under the grant NSC 102-2811-M-009-057 and the Nation Center
for Theoretical Science, Taiwan. S.P. also acknowledges the support from the Alexander von Humboldt Foundation.
DLY was supported by Duke University under the DOE grant DE-FG02-05ER41367 and CYCU under the grant MOST 103-2811-M-033-002.

\section{Appendices}
\subsection{Hall conductivity from the Langevin equation and Boltzmann equations \label{sub:BE}}
In the presence of quasi-particles, we may incorporate the drag force coming from the medium. The equation of motion for the quasi-particles with charge $+1$ then reads
\begin{eqnarray}\label{lorentzf}
\left(\frac{d{\bf p}}{dt}\right)_{R/L}={\bf E}+{\bf v}_{R/L}\times {\bf B}-\xi{\bf p}_{R/L},
\end{eqnarray}
where ${\bf p}$ is the momentum of the quasi-particles and $\xi$ is the drag coefficient. This is basically the Langevin equation in the absence of noise terms. We then take ${\bf v}={j}/j^t$ and ${\bf p}=M{\bf v}$ with $M=M_L=M_R$ being the mass of quasi-particles. We further assume $M\ll T$ such that the chiral symmetry is approximately preserved. Here we also assume that $\xi$ is same for left/right handed particles and isotopic. In the equilibrium state when $d{\bf p}/dt=0$, (\ref{lorentzf}) can be rewritten as
\begin{eqnarray}
E_i=-\epsilon_{ijk}\frac{(j_{R/L})_j}{(j_{R/L})_0}B_k+\xi M\frac{(j_{R/L})_i}{(j_{R/L})_0}.
\end{eqnarray}
By solving the coupled equations for $i=x,y,z$, we find
\begin{eqnarray}
(j_{R/L})_x=0,\quad (j_{R/L})_y=(\sigma_{R/L})_{yy}E_y,\quad (j_{R/L})_z=(\sigma_{R/L})_{yz}E_z,
\end{eqnarray}
where
\begin{eqnarray}\label{eq:sol_Langevin_01}
(\sigma_{R/L})_{yy}=\frac{(j_{R/L})_0}{\xi M\left(1+\frac{B_x^2}{\xi^2M^2}\right)},\quad
(\sigma_{R/L})_{zy}=-\frac{(j_{R/L})_0B_x}{\xi^2 M^2\left(1+\frac{B_z^2}{\xi^2M^2}\right)}.
\end{eqnarray}
One may expect that CME and CSE should lead to non-vanishing $(j_x)_{R/L}$. However, the currents along the magnetic field should deplete in the presence of the drag force, while the currents parallel to the electric field and perpendicular to both the electric and magnetic fields are steady.  

On the other hand, we can express the classical Hall effects via the Boltzmann equations.
In the present of external $\mathbf{E}$ and $\mathbf{B}$ fields,
the Boltzmann equations can be written as, 
\begin{equation}
\frac{df}{dt}=\partial_{t}f+\mathbf{v}\cdot\partial_{x}f-e[\mathbf{v}\cdot\mathbf{E}+\mathbf{v}\times\mathbf{B}]\cdot\frac{\partial}{\partial\mathbf{p}}f=-\frac{f-f_{0}}{\tau},\label{eq:BE_01}
\end{equation}
where \textbf{$\mathbf{v}$} is the velocity of a single particle
with the momentum $\mathbf{p}$, $f(x,p)$ is the distribution function,
$f_{0}$ is $f$ at an equilibrium state. Here we will drop the $R/L$ signs in the derivations for simplicity. In the right handed side,
we use the relaxation time $\tau$ instead of the collision terms.
We can assume the system is very close to an equilibrium state, that
will lead us to expand the $f$ near the $f_{0}$, 
\begin{equation}
f=f_{0}+\delta f,
\end{equation}
with 
\begin{equation}
f_{0}=\frac{1}{e^{(E_{p}-\mu)/T}+1},
\end{equation}
where $E_{p}=|\mathbf{p}|$ is the energy of a massless single particle,$\mu$
is the chemical potential, $T$ is the temperature. Inserting it back
to Eq.(\ref{eq:BE_01}) yields, 
\begin{equation}
\frac{\partial}{\partial t}\delta f+\mathbf{v}\cdot\partial_{x}\delta f-e\left[\mathbf{E}+\mathbf{v}\times\mathbf{B}\right]\cdot\frac{\partial}{\partial\mathbf{p}}\delta f+\mathbf{v}\cdot\left[e\mathbf{E}-\nabla\mu+\frac{E_{p}-\mu}{T}\nabla T\right](-\frac{\partial f_{0}}{\partial E_{p}})=-\frac{\delta f}{\tau},
\end{equation}
For simplicity, we assume the $\delta f(x,p)$, $\mu$ and $T$ are
homogenous in space. In a weak $\mathbf{E}$ field and a strong $\mathbf{B}$
field case, i.e. $\mathbf{E}\ll O(\partial_{x})\ll\mathbf{B}$, we
can also neglect the high order correction $-e\mathbf{E}\cdot\frac{\partial}{\partial\mathbf{p}}\delta f$.
Finally, we get, 
\begin{equation}
\frac{\partial}{\partial t}\delta f-e\mathbf{v}\times\mathbf{B}\cdot\frac{\partial}{\partial\mathbf{p}}\delta f+\mathbf{v}\cdot e\mathbf{E}(-\frac{\partial f_{0}}{\partial E_{p}})=-\frac{\delta f}{\tau}.
\end{equation}
By using the ansatz, $\delta f=\mathbf{p}\cdot\mathbf{G}(E_{p})e^{i\omega t},$
$\mathbf{E}(t)=\mathbf{E}_{0}e^{-i\omega t}$, the Boltzmann equation
can be further simplified as, 
\begin{equation}
(\tau^{-1}-i\omega)\mathbf{p}\cdot\mathbf{G}-e(\mathbf{v}\times\mathbf{B})\cdot\nabla_{p}(\mathbf{p}\cdot\mathbf{G})=e\mathbf{v}\cdot\mathbf{E}_{0}\frac{\partial f_{0}}{\partial E_{p}},
\end{equation}
and the solution is, 
\begin{equation}
\mathbf{G}_{i}=\Gamma_{ji}^{-1}e\mathbf{E}_{0j}\frac{\partial f_{0}}{\partial E_{p}},
\end{equation}
with $\Gamma$ matrix, 
\[
\Gamma_{ij}=(\tau^{-1}-i\omega)\delta_{ij}-\epsilon_{ijk}e\mathbf{B}_{k}.
\]
Then the current induced by the external fields is given by, 
\[
\delta\mathbf{J}_{i}=\int\frac{d^{3}p}{(2\pi)^{3}}\mathbf{v}_{i}\delta f\equiv e\sigma_{ij}\mathbf{E}_{j}(t),
\]
where 
\begin{eqnarray*}
\sigma_{ij} & = & \int\frac{d^{3}p}{(2\pi)^{3}}\mathbf{v}_{i}p_{l}\Gamma_{jl}^{-1}=n\Gamma_{ji}^{-1}.
\end{eqnarray*}
and $n$ is the number density, $n=\frac{1}{3}\int\frac{d^{3}p}{(2\pi)^{3}}f_{0}$.
Note that we have assume $\tau$ as a constant. In the stationary
limit, $\omega\rightarrow0$, if $\mathbf{B}=B\hat{x}$, we get, 
\begin{equation}
\sigma_{zy}=\int\frac{d^{3}p}{(2\pi)^{3}}\mathbf{v}_{i}p_{l}\frac{\partial f_{0}}{\partial E_{p}}\frac{eB\tau^{2}}{E_{p}^{2}+(eB)^{2}\tau^{2}}=\begin{cases}
-\frac{n}{eB},\quad & B\rightarrow\infty,\\
-eI_{10}\tau^{2}B,\quad & B\rightarrow0,
\end{cases}\label{eq:Hall_tensor_BE_01}
\end{equation}
which is consistent with Eq. (\ref{eq:Hall_strong_B_01}) and (\ref{eq:Hall_weak_B_01}),
and 
\begin{equation}
I_{10}=\frac{1}{6\pi^{2}}\int dE_{p}f_{0}(E_{p}),
\end{equation}
is a dimension $1$ quantity. 
           

\end{document}